\documentclass[manuscript,screen]{acmart}
\usepackage{makecell}
\usepackage{multirow}
\usepackage{color}
\usepackage{utfsym}
\usepackage{subfigure}
\usepackage{float}
\usepackage{color}
\usepackage{xcolor}
\usepackage{indentfirst}

\newcommand{\xgy}[1]{{\color{black} #1}}
\newcommand{\revised}[1]{{\color{black} #1}}
\newcommand{\zz}[1]{{\color{black} #1}}
\newcommand{\minor}[1]{{\color{black} #1}}
\AtBeginDocument{%
  \providecommand\BibTeX{{%
    \normalfont B\kern-0.5em{\scshape i\kern-0.25em b}\kern-0.8em\TeX}}}

\setcopyright{acmlicensed}
\acmJournal{CSUR}
\acmYear{2023} \acmVolume{1} \acmNumber{1} \acmArticle{1} \acmMonth{1} \acmPrice{15.00}\acmDOI{10.1145/3625227}
\acmConference[Conference acronym 'XX]{Make sure to enter the correct
  conference title from your rights confirmation emai}{June 03--05,
  2018}{Woodstock, NY}
\acmPrice{15.00}
\acmISBN{978-1-4503-XXXX-X/18/06}

\begin{document}

\title{Cross-Modality Neuroimage Synthesis: A Survey}

\author{Guoyang Xie}
\authornote{The authors contributed equally to this research.}
\email{guoyang.xie@surrey.ac.uk}
\affiliation{%
  \institution{Southern University of Science and Technology}
  \streetaddress{Shenzhen 518055, China}
  \city{Shenzhen}
  \country{China}
  }
\affiliation{%
  \institution{University of Surrey}
  \streetaddress{Guildford GU2 7XH, UK}
  \city{Guildford}
  \country{United Kingdom}}

\author{Yawen Huang}
\authornotemark[1]
\affiliation{%
  \institution{Jarvis Research Center, Tencent YouTu Lab}
  \streetaddress{1 Th{\o}rv{\"a}ld Circle}
  \city{Shenzhen}
  \country{China}}
\email{bear_huang@126.com}

\author{Jinbao Wang}
\authornote{Co-corresponding authors}
\email{linkingring@163.com}
\affiliation{%
  \institution{Southern University of Science and Technology}
  \streetaddress{Shenzhen 518055, China}
  \city{Shenzhen}
  \state{Guangdong}
  \country{China}
  \postcode{518055}
}

\author{Jiayi Lyu}
\affiliation{%
  \institution{University of Chinese Academy of Sciences}
  \city{Beijing}
  \country{China}
}

\author{Feng Zheng}
\authornotemark[2]
\affiliation{%
  \institution{Southern University of Science and Technology}
  \streetaddress{Shenzhen 518055, China}
  \city{Shenzhen}
  \state{Guangdong}
  \country{China}}
  
\author{Yefeng Zheng}
\affiliation{%
  \institution{Jarvis Research Center, Tencent YouTu Lab}
  \city{Shenzhen}
  \country{China}
}

\author{Yaochu Jin}
\affiliation{%
 \institution{Bielefeld University}
 \streetaddress{33619 Bielefeld}
 \city{Bielefeld}
 \country{Germany}}
\affiliation{%
  \institution{University of Surrey}
  \streetaddress{Guildford GU2 7XH, UK}
  \city{Guildford}
  \country{United Kingdom}}

\renewcommand{\shortauthors}{Guoyang, Jinbao and Yawen, et al.}

\begin{abstract}
Multi-modality imaging improves disease diagnosis and reveals distinct deviations in tissues with anatomical properties. The existence of completely aligned and paired multi-modality neuroimaging data has proved its effectiveness in brain research. However, collecting fully aligned and paired data is expensive or even impractical, since it faces many difficulties, including high cost, long acquisition time, image corruption, and privacy issues. An alternative solution is to explore unsupervised or weakly supervised learning methods to synthesize the absent neuroimaging data. In this paper, we provide a comprehensive review of cross-modality synthesis for neuroimages, from the perspectives of weakly supervised and unsupervised settings, loss functions, evaluation metrics, imaging modalities, datasets, and downstream applications based on synthesis. We begin by highlighting several opening challenges for cross-modality neuroimage synthesis. Then, we discuss representative architectures of cross-modality synthesis methods under different supervisions. This is followed by a stepwise in-depth analysis to evaluate how cross-modality neuroimage synthesis improves the performance of its downstream tasks. Finally, we summarize the existing research findings and point out future research directions. All resources are available at \href{https://github.com/M-3LAB/awesome-multimodal-brain-image-systhesis}{https://github.com/M-3LAB/awesome-multimodal-brain-image-systhesis}.
\end{abstract}



\maketitle


\section{Introduction}
\begin{figure}[t]
    \centering
    \includegraphics[width=1\linewidth]{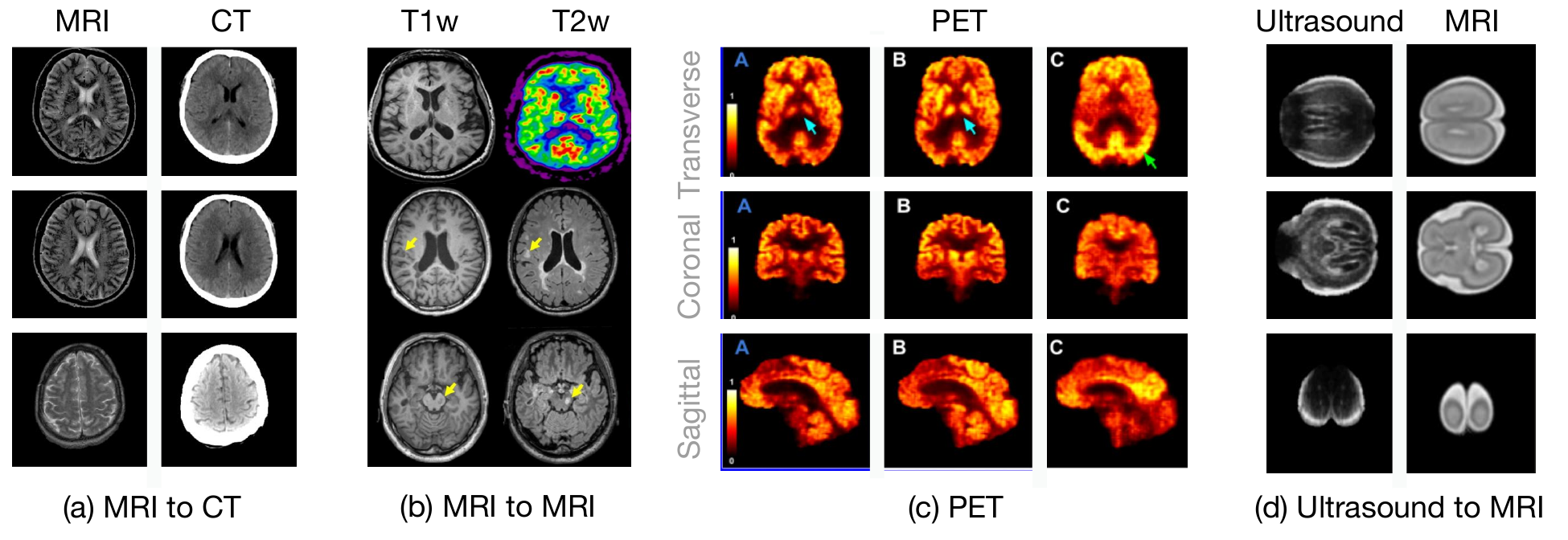}
    \caption{Multi-modality synthesis. (a) MRI to CT~\cite{Han2017MRbasedSC}, (b) MRI to PET~\cite{Zhang202118FflorbetapirPF}, (c) PET~\cite{Zhou2021SynthesizingMP}, and (d) ultrasound to MRI~\cite{Jiao2020SelfSupervisedUT}.}
    \label{fig:synthesis_range}
\end{figure}

\xgy{\textbf{The necessity of cross-modality neuroimage synthesis.}} The majority of multi-center neuroimaging datasets~\cite{Aljabar2011ACM,Siegel2019CancerS2} are often high-dimensional and heterogeneous, which is shown in Fig.~\ref{fig:synthesis_range}. For example, positron emission tomography (PET) and magnetic resonance imaging (MRI) are classic medical imaging techniques that provide detailed anatomic and physiological images of different organs for auxiliary diagnosis or monitoring treatments. The paired or registered multi-modality data provide more complementary information to investigate certain pathologies, including neurodegeneration. However, it is not feasible to acquire a full set of wholly paired and aligned multi-modality neuroimaging data, considering that: 
\begin{itemize}
    \item Collecting multi-modality neuroimaging data is very costly. For example, a normal MRI can take more than one thousand dollars in some countries;
    \item Many medical institutions cannot share their data since medical data are predominantly restricted by local regulations, despite that identifiable information can be removed to protect the privacy of patients;
    \item Patient motions may result in severely misaligned neuroimaging data.
\end{itemize}
As a result, there is a clear need to handle the absent data through a cross-modality synthesis method.

\begin{figure}[h]
\centering
\subfigure[Statistical Result]{
    \label{fig:statistical_detail_result}
    \includegraphics[width=0.3\textwidth, height=0.2\textwidth]{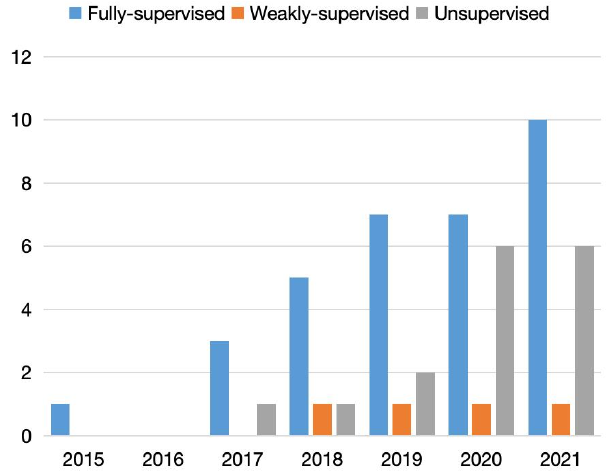}}
    \hspace{4pt}
\subfigure[Modality Synthesis Range]{
    \label{fig:statistical_synthesis_range}
    \includegraphics[width=0.3\textwidth, height=0.2\textwidth]{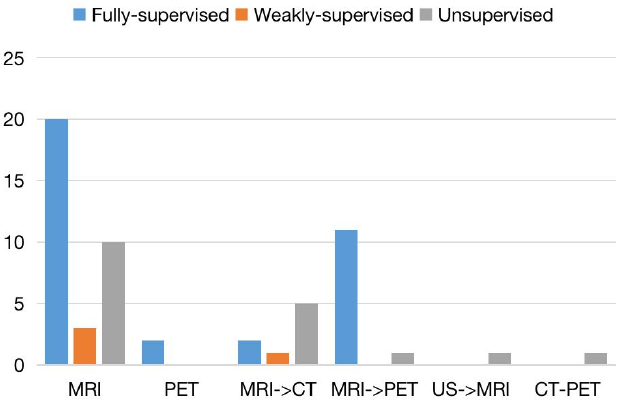}}
    \hspace{4pt}
\subfigure[Downstream Task]{
    \label{fig:statistical_downstream_task}
    \includegraphics[width=0.3\textwidth, height=0.2\textwidth]{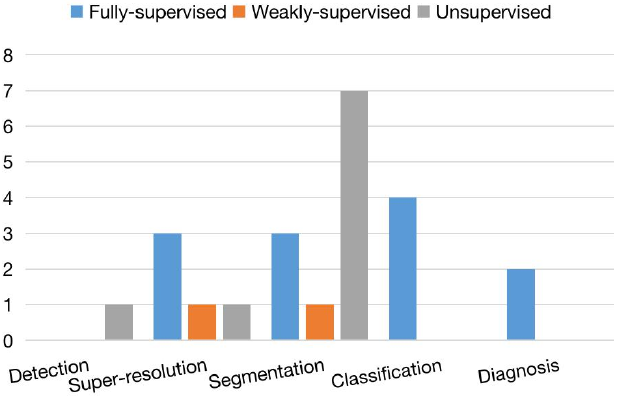}}
    \caption{Trend of each learning manner in multi-modality brain image synthesis literature. (a) The number of papers published chronologically for different levels of supervision papers. (b) The distribution of modality synthesis mode according to the level of supervision. (c) The number of various downstream tasks according to the level of supervision.}
\label{fig:trend_diff_learning}
\end{figure}

\xgy{\textbf{Cross-modality Generative Adversarial Networks (GANs) in medical imaging community.} Previously, the key problems in the medical imaging community were: (1) how to fasten multi-contrast MRI reconstruction, (2) how to enhance the image quality of MRI or computed tomography (CT) scan, (3) medical image registration, and (4) fine-grained medical image segmentation. Experienced researchers can quickly iteratively construct innovative algorithms because the majority of them have rather mature answers, ensuring feasibility and high precision from design to product deployment. Cross-modality neuroimage synthesis algorithms, however, were still in their infancy as of 2018 since the synthesized neuroimage quality could not satisfy radiologist's demands. The situation changed substantially when CycleGAN~\cite{Zhu2017UnpairedIT} emerged. As shown in Fig.~\ref{fig:trend_diff_learning}(a), we can easily observe that the number of unsupervised and weakly supervised learning methods is increasing. The researchers have paid more attention to unsupervised and weakly supervised learning methods. Fig.~\ref{fig:trend_diff_learning}(c) indicates the number of downstream tasks with each supervision level. It can be easily observed that most unsupervised and weakly supervised methods are jointly optimized with the segmentation task. However, classification and diagnosis tasks are ignored in unsupervised learning and weakly supervised learning methods, to which we believe future work should pay more attention. Fig.~\ref{fig:trend_diff_learning}(b) presents modality synthesis according to the level of supervision. We notice that most of the algorithms conduct cross-modality synthesis for MRI. However, PET and MRI have not received enough attention in unsupervised learning and weakly supervised learning algorithms. We expect future work to propose a uniform generator to synthesize an arbitrary modality among PET, MRI or CT in an unsupervised or weakly supervised learning manner.}


\begin{figure}[htbp]
    \centering
    \includegraphics[width=0.75\linewidth]{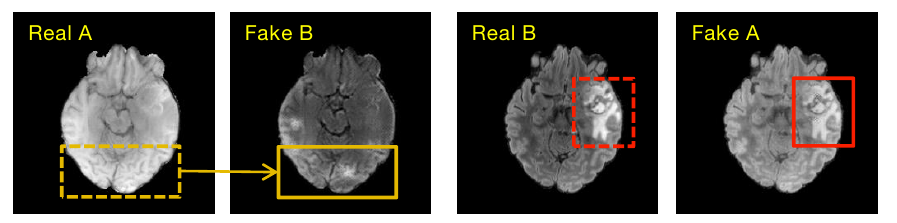}
    \caption{A failed case in multi-modality brain image synthesis. In addition to generating low-resolution images, another problem is that the disease-related regions cannot be synthesized well. For example, when the target modality (Fake B) is generated from the input real modality (Real A), there exist failed regions (box) that are learned from the original ones (dashed box).}
    \label{fig:existing_dismerit}
\end{figure}

\xgy{\textbf{Look back cross-modality neuroimage synthesis.}} 
\xgy{From the evolution standpoint, natural image-to-image translation \revised{leads} the development of cross-modality neuroimage synthesis. Inspired by \textit{dictionary learning}~\cite{Aharon2005KSVDA}, Roy~\textit{et al.}~\cite{Roy2013MagneticRI} trained two dictionaries where the input image is used to find similar patches in a source modality dictionary and the corresponding target modality counterpart will be extracted from the target dictionary to generate the desirable modality data. The work of Huang~\textit{et al.}~\cite{Huang2016GeometryRJ} \revised{improves} the quality of cross-modality synthesis by imposing a graph Laplacian constraint in a joint dictionary learning framework. Wang~\textit{et al.}~\cite{Wang2017RegionEnhancedJD} synthesizes the missing DT images from T1-w scans by learning a region-enhanced joint dictionary. When \textit{pix2pix}~\cite{Isola2017ImagetoImageTW} \revised{was} released in 2017, the situation drastically changed. Most supervised cross-modality neuroimage synthesis algorithms adopt variants of pix2pix. Maspero~\textit{et al.}~\cite{Maspero2018DoseEO} directly employ pix2pix to synthesize CT scans from MRI. The work of Olut~\textit{et al.}~\cite{Olut2018GenerativeAT} synthesizes magnetic resonance angiography (MRA) from T1 and T2 with the addition of a steerable filter loss on pix2pix. Furthermore, \textit{CycleGAN}~\cite{Zhu2017UnpairedIT} boosts the performance of unsupervised cross-modality neuroimage synthesis. Hiasa~\textit{et al.}~\cite{Hiasa2018CrossmodalityIS} employ gradient consistency loss to optimize the edge map of the synthesized neuroimage. Zhang~\textit{et al.}~\cite{Zhang2018TranslatingAS} propose two segmentation networks to segment the corresponding image modality into semantic labels and provide implicit shape constraints on the anatomy during translation. Chen~\textit{et al.}~\cite{Chen2018SemanticAwareGA} propose a similar method to the work of Zhang~\textit{et al.}~\cite{Zhang2018TranslatingAS}. The only difference between them is that the segmentation network of Chen~\textit{et al.}~\cite{Chen2018SemanticAwareGA} is trained offline and fixed during the training phase of the image translation network. \textit{\textbf{The inherent properties of medical images are ignored even though natural image-to-image translation approaches reveal many insights into cross-modality neuroimage synthesis.}} For instance, Fig.~\ref{fig:existing_dismerit} shows a failed case of CycleGAN~\cite{Zhu2017UnpairedIT} for cross-modality neuroimage synthesis. In particular, the lesion region of the target modality (red box of Fake A) cannot be accurately synthesized in comparison to the lesion region of the source modality (red dashed box of Real B), particularly for the structural details. Therefore, it is necessary to investigate more deeply to characterize the network architecture for the imaging principle of neuroimages.}

\revised{\textbf{\textit{Motivation and objectives.}} In recent years, unsupervised cross-modality synthesis has garnered a growing amount of attention, and a number of practical solutions have been proposed to tackle this complex problem. There is no systematic review of the current state of cross-modality neuroimage synthesis, particularly for unsupervised methods and their application in downstream tasks. The primary objective of this survey is to provide a comprehensive study on cross-modality neuroimage synthesis and a solid foundation for newcomers interested in this field and related topics. This paper offers a comprehensive summary of the state-of-the-art cross-modality neuroimage synthesis methods. In addition, we investigate in depth how to use cross-modality neuroimage synthesis for tasks such as tumor segmentation, lesion detection, and diagnosis. Finally, the remaining open questions for future research are discussed.} 

\revised{\textbf{\textit{Comparison to existing surveys. }} Table~\ref{tab:survey_comparison} compares the present survey with two existing ones concerning the level of supervision, the range of modality synthesis, and the downstream task. The work of~\cite{Yi2019GenerativeAN} provides a comprehensive review of generative adversarial networks (GANs) in medical imaging before 2019, including single-modality synthesis, cross-modality synthesis, and the usage of the GAN in different downstream tasks, e.g., classification, segmentation, registration. However, this work is different from previous reviews~\cite{Yi2019GenerativeAN,Zhao2020DeepLO} in three aspects: (1) The work~\cite{Yi2019GenerativeAN} was published in 2019, where most of the cross-modality synthesis methods reviewed~\cite{Yi2019GenerativeAN} are supervised. In other words, most synthesis algorithms require fully paired medical data for training. (2) The performance of the downstream tasks when leveraging the synthesized results is missing in the review, which is of great importance since the final purpose of cross-modality synthesis serves as an auxiliary procedure for the downstream tasks, e.g., segmentation. Another work in~\cite{Zhao2020DeepLO} comprehensively reviews the state-of-the-art deep learning-based methods applied to brain MRI, including segmentation, registration, and diagnosis. It touches upon cross-modality brain synthesis but does not go into detail. Furthermore, a review of the taxonomy is presented in~\cite{Zhao2020DeepLO}, which focuses on the task without considering the level of supervision. In addition, it reviews brain magnetic resonance imaging-related work, mainly ignoring other common imaging modalities such as CT and PET. (3) Our review offers a comprehensive analysis and in-depth understanding of the loss function and evaluation metrics. The loss function is classified into three streams based on its function in cross-modality neuroimage synthesis. Furthermore, our study extensively examines the limitations of current metrics and provides insight into the future route to evaluate the quality of synthesized neuroimages.}

\revised{
\begin{table}[]
\renewcommand{\arraystretch}{1.3}
\caption{\revised{Comparison on related surveys in terms of supervision level, modality synthesis mode and downstream tasks.}}
\centering
\resizebox{\linewidth}{!}{
\begin{tabular}{c|p{1.1cm}<{\centering}p{1.1cm}<{\centering}p{1.1cm}<{\centering}|ccccc|cccc}

\hline
\multirow{2}{*}{\textbf{Survey}}      & \multicolumn{3}{c|}{\textbf{Supervison Level}}     & \multicolumn{5}{c|}{\textbf{Modality Synthesis Mode}}    & \multicolumn{4}{c}{\textbf{Downstream Tasks}}                                                                         \\ \cline{2-13} 
& Fully-                   & weakly-              & Un-                 & MRI$\rightarrow$MRI        & PET$\rightarrow$PET        & MRI$\rightarrow$CT         & US$\rightarrow$MRI         & CT$\rightarrow$PET         & Segmentation                 & Classification               & Registration                 & Diagnosis                    \\ \hline
Yi~\textit{et al.}~\cite{Yi2019GenerativeAN} & $\usym{1F5F8}$ &                              &                              & $\usym{1F5F8}$ & $\usym{1F5F8}$ & $\usym{1F5F8}$ &                              & $\usym{1F5F8}$ &                              &                              &                              &                              \\
Zhao~\textit{et al.}~\cite{Zhao2020DeepLO}   & $\usym{1F5F8}$ &                              &                              & $\usym{1F5F8}$ &                              &                              &                              &                              & $\usym{1F5F8}$ & $\usym{1F5F8}$ & $\usym{1F5F8}$ & $\usym{1F5F8}$ \\
\textbf{Ours }      & $\usym{1F5F8}$ & $\usym{1F5F8}$ & $\usym{1F5F8}$ & $\usym{1F5F8}$ & $\usym{1F5F8}$ & $\usym{1F5F8}$ & $\usym{1F5F8}$ & $\usym{1F5F8}$ & $\usym{1F5F8}$ & $\usym{1F5F8}$ & $\usym{1F5F8}$ & $\usym{1F5F8}$ \\ \hline
\end{tabular}
}\label{tab:survey_comparison}
\end{table}
}

\xgy{\textit{\textbf{Contributions.}}} \xgy{The main contributions of this survey paper can be summarized as follows: 
\begin{itemize}
    \item To the best of our knowledge, it is the first work to provide an in-depth review of cross-modality brain image synthesis by considering the level of supervision, especially for both unsupervised and weakly supervised cross-modality synthesis.  
    \item It provides a comprehensive review of the relationship between cross-modality synthesis and its downstream tasks. This survey focuses on how to synthesize appropriate cross-modality brain images to improve downstream tasks, such as image segmentation, registration, and diagnosis.
    \item It summarizes the main issues and potential challenges in cross-modality brain image synthesis, highlighting the underlying research directions for future work. 
\end{itemize}}

\textbf{Organization.} 
The rest of the paper is organized as follows. We begin with a chronological overview of cross-modality brain image synthesis and review these methods in Section~\ref{sec:chronological}. In Section~\ref{sec:learn_paradigm}, we review cross-modality neuroimage synthesis on the basis of the level of supervision. Next, we review recent advances in modality synthesis in Section~\ref{sec:range}. In Section~\ref{sec:down_task}, we provide an analysis of how cross-modality synthesis significantly improves the performance of the downstream tasks. Then, in Section~\ref{sec:dataset}, we describe the popular datasets and go over various loss functions in Section~\ref{sec:losses} and take a retrospective view of the evaluation metrics in Section~\ref{sec:metrics}. In the end, we provide future research directions for brain neuroimage cross-modality synthesis in Section~\ref{sec:future_work}.

\section{Chronological Review}\label{sec:chronological}
\begin{figure*}[t]
    \centering
    \includegraphics[width=1\linewidth]{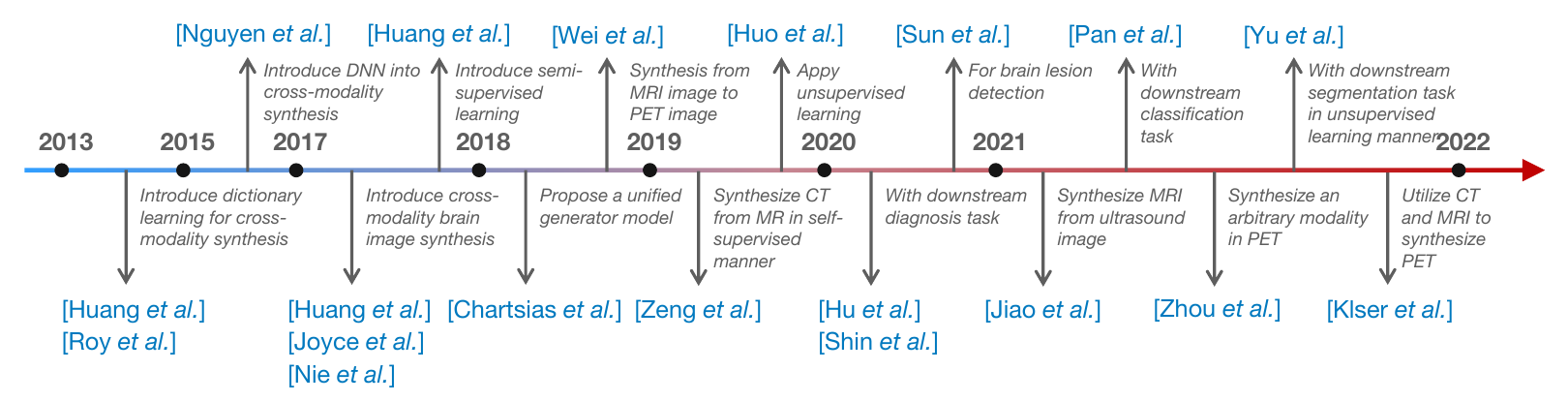}
    \caption{A chronological review of multi-modality brain image synthesis.}
    \label{fig:timeline}
\end{figure*}

Fig.~\ref{fig:timeline} gives a chronological overview of the cross-modality brain synthesis methods according to the level of supervision, the relevant downstream tasks, and the mode of modality synthesis. \xgy{In 2013, Roy~\textit{et al.}~\cite{Roy2013MagneticRI} was the first to introduce the dictionary method~\cite{yang2010image} into cross-modality neuroimage synthesis. Ye~\textit{et al.}~\cite{Ye2013ModalityPC} propose a modality propagation method and prove that the proposed model can be derived from the generalization of label propagation strategy~\cite{Heckemann2006AutomaticAB}, and show applications to arbitrary modality synthesis. In 2015, Nguyen~\textit{et al.}~\cite{Nguyen2015CrossDomainSO} trained a location-sensitive deep network to integrate intensity features and spatial information, more accurately synthesizing the results for the same problem posed by~\cite{Ye2013ModalityPC}}. In 2017, the works in~\cite{Huang2017DOTEDC, Joyce2017RobustMM, Nie2017MedicalIS} introduced cross-modality brain image synthesis into the medical GAN community. Their methods are supervised, i.e., their training data are fully paired. Huang~\textit{et al.}~\cite{Huang2017DOTEDC} construct a closed loop filter learning strategy to learn the convolutional sparse coding (CSC), which can eliminate the requirement of large-scale training data. Meanwhile, it is also the first to undertake super-resolution and multi-modality neuroimaging data in MRI. The authors of~\cite{Joyce2017RobustMM} propose a multi-modality invariant latent embedding model for synthesis. The purpose of this method is to utilize the mutual information from multi-modality maximally and fuse them into the generated image. Nie~\textit{et al.}~\cite{Nie2017MedicalIS} introduce a synthesis method by translating brain MRI data to CT. The authors incorporate the detailed information from brain MRI into the GAN model to generate the brain CT. After that, Huang~\textit{et al.}~\cite{Huang2017SimultaneousSA} provide a weakly supervised learning approach to cross-modality brain image synthesis. The work in~\cite{Huang2017SimultaneousSA} regards the unpaired data as auxiliary resources. They propose a hetero-domain image alignment method to enforce the correspondence for unpaired auxiliary data, which can directly substantiate the benefits of the combination with a few paired data and massive unpaired data. Chartsias~\textit{et al.}~\cite{Chartsias2018MultimodalMS} firstly propose a unified generator model for various MRI modalities. Huo~\textit{et al.}~\cite{Huo2019SynSegNetSS} firstly apply an unsupervised learning method to cross-modality brain image synthesis. In other words, the multi-modality training data are unpaired. Specifically, they adopt CycleGAN~\cite{Zhu2017UnpairedIT} to generate the target modality from the source modality. 

After that, the authors in~\cite{Huo2019SynSegNetSS} leverage both the synthesized and source modality data for segmentation. It is also the first to employ unsupervised learning methods for downstream tasks. Wei~\textit{et al.}~\cite{Wei2018LearningMC} provide a complex synthesis approach to synthesize from MRI image to PET image. Specifically, Sketcher-Refiner GANs proposed by~\cite{Wei2018LearningMC} decompose the synthesis problem as a sketch-refinement process, in which the sketchers generate the preliminary anatomical and physiological information, and the refiner refines the structure of tissue myelin content. Pan~\textit{et al.}~\cite{Pan2021DiseaseimagespecificLF} provide a method to optimize the cross-modality synthesis and diagnosis tasks jointly. They design a disease-image-specific network (DSNet) by feeding the features generated from the disease-image-specific network into feature-consistency GANs (FC-GANs) to generate the target domain neuroimaging data. Since DSNet is closely associated with FC-GANs, the missing target domain data can be synthesized in a diagnosis-oriented manner. Hu~\textit{et al.}~\cite{Hu2020BrainMT} and Shin~\textit{et al.}~\cite{Shin2020GANDALFGA} are the first to utilize the synthesized neuroimaging data to improve the performance of a classification task. Zhou~\textit{et al.}~\cite{Zhou2021SynthesizingMP} propose a generator to synthesize an arbitrary modality in PET. Klser~\textit{et al.}~\cite{Klser2021ImitationLF} utilize two modalities of data, i.e., CT and MRI, to synthesize PET data. Yu~\textit{et al.}~\cite{Yu2021MouseGANGM} jointly optimize the synthesis and segmentation problems using unsupervised learning methods. Jiao~\textit{et al.}~\cite{Jiao2020SelfSupervisedUT} synthesize MRI from ultrasound image using a new fusion scheme to utilize various modalities from unpaired data. Zeng~\textit{et al.}~\cite{Zeng2019HybridGA} synthesize CT from MR by using self-supervised methods. 

\begin{table}[t]
\caption{A brief summary of cross-modality brain image synthesis methods across different MRI contrast or from MRI to CT. Note that there are three levels of supervision (Sup.), i.e., fully supervised (\textcolor{gray}{F}), weakly supervised (\textcolor{blue}{W}) and unsupervised (\textcolor{red}{U}). The downstream tasks (DST) contains super-resolution (SR), segmentation (S),  detection (D), inpainting (I) and registration (R). }
\renewcommand{\arraystretch}{1.1}
\resizebox{\textwidth}{!}{
    \begin{tabular}{llp{1cm}p{4.5cm}p{3.4cm}p{4cm}lp{1cm}}
    \hline
    \textbf{Publication} & \textbf{Name/Arch.} & \textbf{Sup.}& \textbf{Loss/Method} & \textbf{Dataset} & \textbf{Metrics} & \textbf{Align} & \textbf{DST} 
    \\\hline
    \multicolumn{8}{c}{\textit{MRI $\rightarrow$ MRI}} \\\hline
        Huang \textit{et al.} \cite{Huang2019SimultaneousSA}& WEENIE & \textcolor{blue}{W} & Sparse coding &  IXI, NAMIC & PSNR, SSIM & $\usym{1F5F8}$  & SR \\ 
        Huang \textit{et al.} \cite{Huang2017DOTEDC} & DOTE & \textcolor{gray}{F}  & Sparse coding & IXI, NAMIC & PSNR, SSIM &  $\usym{1F5F8}$ & SR \\
        Huang \textit{et al.} \cite{Huang2019CoCaGANCC} & CoCa-GAN & \textcolor{gray}{F} & $L_{gan}$, $L_{edge}$, $L_{gdl}$, $L_{tumor}$ & BraTS15 & PSNR, NMSE, SSIM & $\usym{1F5F8}$ &  \\
        Huang \textit{et al.} \cite{Huang2020MCMTGANMC} & MCMT-GAN & \textcolor{gray}{F} & $L_{gan}$, $L_{cyc}$, $L_{mani}$, $L_{fault}$  &  IXI, NAMIC & PSNR, SSIM, DSC & $\usym{1F5F8}$ & S \\
        Huang \textit{et al.}  \cite{Huang2018CrossModalityIS}  & WAG & \textcolor{blue}{W} & \makecell[l]{$L_{spc}$, $L_{mmd}$, $L_{gcr}$} & IXI, NAMIC & PSNR, SSIM & $\usym{1F5F8}$ &  \\
        Huang \textit{et al.} \cite{Huang2020SuperResolutionAI} & GAN & \textcolor{red}{U} & \makecell[l]{$L_{gan}$, $L_{cont}$, $L_{text}$, $L_{cyc}$} & IXI, HCP & PSNR, SSIM & $\usym{2717}$ & SR, I \\
        He \textit{et al.} \cite{He2021AutoencoderBS} & \href{https://github.com/YufanHe/self-domain-adapted-network}{Autoencoder} & \textcolor{red}{U} & $L_{ae}$, $L_{oth}$   & IXI & MSE, SSIM & $\usym{2717}$ & \\
        Yang \textit{et al.} \cite{Yang2021AUH} & Hyper-GAN & \textcolor{red}{U} & \makecell[l]{$L_{gan}$, $L_{cyc}$, $L_{rec1}$, $L_{iden}$, $L_{cla}$} & IXI, BraTS19 & PSAR, SSIM, MAE & $\usym{2717}$ &  \\
        Sun \textit{et al.} \cite{Sun2020AnAL} & ANT-GAN & \textcolor{red}{U} & $L_{gan}$, $L_{cyc}$, $L_{am}$ & BraTS18 & P, R, F1S, VS & $\usym{2717}$ &  S, D  \\
        Guo \textit{et al.} \cite{Guo2021AnatomicAM} & \href{https://github.com/guopengf/CG-SAMR}{CG-SAMR} & \textcolor{red}{U} & \makecell[l]{$L_{gan}$, $L_{fm}$, $L_{sc}$, $L_{cm}$, $L_{rec1}$, $L_{cyc}$} & D5 & DSC & $\usym{2717}$ & S  \\
        Shen \textit{et al.} \cite{Shen2021MultiDomainIC} & ReMIC & \textcolor{blue}{W} & \makecell[l]{$L_{gan}$, $L_{cyc}$, $L_{rec1}$, $L_{dice}$} & BraTS18, ProstateX & NRMSE, PSNR, SSIM & $\usym{2717}$ & S, I  \\
        Tomar \textit{et al.} \cite{Tomar2021SelfAttentiveSA} & SASAN & \textcolor{red}{U} & \makecell[l]{$L_{sgan}$, $L_{cyc}$, $L_{iden}$, $L_{reg}$, $L_{dice}$} & BraTS15 & DSC, ASSD &  $\usym{2717}$ & S \\ 
        Charsias \textit{et al.} \cite{Chartsias2018MultimodalMS} & Autoencoder & \textcolor{gray}{F} & $L_{ae}$ & ISLES15, BraTS15, IXI & MSE, PSNR, SSIM & $\usym{1F5F8}$  & \\
        Chen \textit{et al} \cite{Chen2021ABCnetAB} & \href{https://github.com/cljun27/ABCnet}{ABCNet}  & \textcolor{gray}{F} & \makecell[l]{$L_{gan}$, $L_{sm}$, $L_{liu}$, $L_{cor}$} & BCP, dHCP, HCP & DSC, ASSD, CC, CIV, HD95 & $\usym{1F5F8}$  &   \\
       Bône \textit{et al.} \cite{Bne2021ContrastEnhancedBM} & Autoencoder & \textcolor{gray}{F} & $L_{rec2}$ & -- &  PSNR, SSIM, L2, t-L2 & &  \\
       Dar \textit{et al.} \cite{Dar2019ImageSI}  & pGAN & \textcolor{gray}{F} & $L_{cgan}$, $L_{p2p}$, $L_{prec}$, $L_{cyc}$  & MIDAS, IXI, BraTS15 & PSNR, SSIM & $\usym{2717}$ &  \\
       Jog \textit{et al.} \cite{Jog2017RandomFR} & REPLICA & \textcolor{gray}{F} & Random forest & MMRR & PSNR, SSIM, UQI & $\usym{1F5F8}$  &  \\
       Joyce \textit{et al.} \cite{Joyce2017RobustMM} & Autoencoder & \textcolor{gray}{F} & $L_{p2p}$ &ISLES15, BraTS15 & MSE &  $\usym{2717}$ &  \\
       Kwon \textit{et al.} \cite{Kwon2019GenerationO3} & \href{https://github.com/cyclomon/3dbraingen}{Autoencoder} & \textcolor{gray}{F} & $L_{gan}$ & ANDI, BraTS18, ATLAS & MMD, MSSSIM & $\usym{1F5F8}$ &  \\
       Qu \textit{et al.} \cite{qu2020multimodal} & MCGAN & \textcolor{red}{U} & $L_{gan}$, $L_{ssim}$, $L_{rec2}$, $L_{cyc}$, $L_{edge}$ & BraTS15 & PSNR, SSIM, DSC & $\usym{1F5F8}$  & S \\
       Lee \textit{et al.} \cite{Lee2019CollaGANCG} & CollaGAN & \textcolor{gray}{F} &  $L_{gan}$, $L_{ssim}$ & D9 & NMSE, SSIM &  $\usym{1F5F8}$  &  \\
       Li \textit{et al.} \cite{Li2019DiamondGANUM} & \href{https://github.com/hongweilibran/DiamondGAN}{DiamondGAN} & \textcolor{gray}{F} & $L_{rec1}$, $L_{gan}$ & D10 &PSNR, MAE &  $\usym{2717}$ &  \\
       Liu \textit{et al.} \cite{Liu2021AUC}  & Conditional GAN & \textcolor{gray}{F} & $L_{gan}$, $L_{p2p}$, $L_{mc}$ & BraTS18   & L1, SSIM, PSNR, IS & $\usym{1F5F8}$  & S \\
       Sharma \textit{et al.} \cite{Sharma2020MissingMP} & MM-GAN & \textcolor{gray}{F} & $L_{lsgan}$, $L_{p2p}$ & ISLES15, BraTS18  & MSE, PSNR, SSIM & $\usym{1F5F8}$ &  \\
       Xin \textit{et al.} \cite{Xin2020MultiModalityGA} & \href{https://github.com/hellopipu/TC-MGAN}{TC-MGAN} & \textcolor{gray}{F} & $L_{gan}$, $L_{p2p}$, $L_{cls}$, $L_{seg1}$ & BraTS18 & PSNR, SSIM & $\usym{1F5F8}$  &  \\
       Yang \textit{et al.} \cite{yang2018mri}  \cite{yang2020mri} & cGANs & \textcolor{red}{U} & $L_{cgan}$, $L_{rec1}$  & MRBrainS13, ANDI, RIRE, iSeg17, BraTS15 & MAE, PSNR, MI, F-score, DSC & $\usym{1F5F8}$  & S, R \\
       Yu \textit{et al.} \cite{Yu2020SampleAdaptiveGL} & SA-GAN & \textcolor{gray}{F} & $L_{gan}$, $L_{iden}$  & BraTS15, SSIS15 & PSNR, NMSE, SSIM & $\usym{1F5F8}$ &  \\
       Yu \textit{et al.} \cite{Yu20183DCB} & 3D cGAN & \textcolor{gray}{F} & $L_{cgan}$, $L_{iden}$ & BraTS15 & PSNR, NMSE, DSC & $\usym{1F5F8}$ &  S\\
       Yurt \textit{et al.} \cite{Yurt2021mustGANMG} & \href{https://github.com/icon-lab/mrirecon}{mustGAN} &  \textcolor{gray}{F} & $L_{cgan}$, $L_{iden}$ & IXI, ISLES15 & PSNR, SSIM & $\usym{1F5F8}$ &  \\
       Zhou \textit{et al.} \cite{Zhou2020HiNetHN}  & Hi-Net &  \textcolor{gray}{F} & $L_{cgan}$, $L_{rec1}$ & BraTS18, ISLES15 &  PSNR, SSIM, NMSE & $\usym{1F5F8}$  & \\
       Zuo \textit{et al.} \cite{Zuo2021DMCFusionDM} & DMC-Fusion &\textcolor{gray}{F} & $L_{rec1}$, $L_{ssim}$ & D15 & MI, Q, FMI, PSNR, SSIM & $\usym{1F5F8}$  &  \\
       Bian \textit{et al.} \cite{bian2022dda} & DDA-Net & \textcolor{red}{U} & $L_{gan}$, $L_{cyc}$, $L_{seg}$ & MRBrainS18 & DSC, SEN, SPE, ROC & $\usym{1F5F8}$  & S \\
       \hline
           
    \multicolumn{8}{c}{\textit{MRI $\rightarrow$ CT}} \\\hline
        Hemsley \textit{et al.} \cite{Hemsley2020DeepGM}  & cGAN  & \textcolor{blue}{W} & $L_{cgan}$, $L_{p2p}$  & D1 & MAE & $\usym{1F5F8}$ &  \\ 
        Huo \textit{et al.} \cite{Huo2019SynSegNetSS} & \href{https://github.com/MASILab/SynSeg-Net}{SynSeg-Net} & \textcolor{red}{U} & \makecell[l]{$L_{gan}$, $L_{cyc}$, $L_{seg}$}  & D3 & DSC, ASD & $\usym{2717}$ & S  \\
        Yang \textit{et al.} \cite{Yang2020UnsupervisedMS} & sc-cycleGAN  & \textcolor{red}{U} & $L_{lsgan}$, $L_{cyc}$ & D6 & MAE, PSNR, SSIM &  $\usym{2717}$ & \\ 
        Zeng \textit{et al.} \cite{Zeng2019HybridGA} & 2D-cGAN & \textcolor{red}{U} & \makecell[l]{$L_{sgan}$, $L_{cyc}$, $L_{rec1}$} & D7 & MAE, PSNR &  $\usym{2717}$ & \\
        Klser \textit{et al.} \cite{Klser2021ImitationLF} & pCT & \textcolor{red}{U} & $L_{rec2}$ & D8 & MAE & $\usym{1F5F8}$ &  \\        
        Nie \textit{et al.} \cite{Nie2017MedicalIS} & GAN & \textcolor{red}{U} &$L_{gan}$, $L_{gdl}$ & ANDI & PSNR, MAE & $\usym{1F5F8}$ &  \\
        Zuo \textit{et al.} \cite{Zuo2021DMCFusionDM} & DMC-Fusion & \textcolor{gray}{F} & $L_{rec1}$, $L_{ssim}$ & D15 & MI, Q, FMI, PSNR, SSIM  & $\usym{1F5F8}$  &  \\  
        Huynh \textit{et al.} \cite{huynh2015estimating} &  -- & \textcolor{gray}{F} & Random forest & ANDI & MAE, PSNR & $\usym{1F5F8}$  &  \\  
    
    \hline
    \end{tabular}
}
\label{tab:summary-methods-1}
\end{table}

\begin{table}[th]
\caption{A brief summary of cross-modality brain image synthesis methods for PET synthesis and US $\rightarrow$ MRI. Note that there are three levels of supervision (Sup.), i.e., fully supervised (\textcolor{gray}{F}), weakly supervised (\textcolor{blue}{W}) and unsupervised (\textcolor{red}{U}) manner. Downstream task (DST) contains super-resolution (SR), classification (C), detection (D), inpainting (I), registration (R), explainability (E), and diagnosis (Diag).}
\renewcommand{\arraystretch}{1.1}
\resizebox{\textwidth}{!}{
    \begin{tabular}{llp{1cm}p{4.5cm}p{3.4cm}p{4cm}lp{1cm}}
    \hline
    \textbf{Publication} & \textbf{Name/Arch.} & \textbf{Sup.}& \textbf{Loss/Method} & \textbf{Dataset} & \textbf{Metrics} & \textbf{Align} & \textbf{DST} 
    \\\hline
    \multicolumn{8}{c}{\textit{MRI $\rightarrow$ PET}} \\\hline
        Wang \textit{et al.} \cite{Wang2018LocalityAM} & LA-GANs & \textcolor{gray}{F} & $L_{cgan}$, $L_{p2p}$ & D2 & PSNR & $\usym{1F5F8}$ & SR \\ 
        Hu \textit{et al.} \cite{Hu2020BrainMT} & Bidirectional GAN & \textcolor{gray}{F} & $L_{gan}$, $L_{rec1}$, $ L_{prec}$, $L_{kl}$ & ADNI & PSNR, SSIM  & $\usym{1F5F8}$ &  \\
        Hu \textit{et al.} \cite{Hu2022BidirectionalMG} & BMGAN & \textcolor{gray}{F} &  $L_{gan}$, $L_{rec1}$, $ L_{prec}$, $L_{kl}$ & ADNI & MAE, PSNR, MSSSIM, FID & $\usym{1F5F8}$  &  \\
        Liu \textit{et al.} \cite{Liu2022AssessingCP} & \href{https://github.com/Candyeeee/JSRL}{JSRL}  & \textcolor{gray}{F} &  $L_{gan}$, $L_{rec1}$, $L_{cls}$ & CLAS, ADNI, AIBL & AUC, BAC, SEN, SPE, F1S &  $\usym{1F5F8}$ &  C \\
        Pan \textit{et al.} \cite{Pan2018SynthesizingMP} & LM$^3$IL & \textcolor{gray}{F}  & $L_{gan}$, $L_{cyc}$ & ADNI & ACC, SEN, SPE, F1S, MCC, AUC & $\usym{1F5F8}$ &  C \\
        Pan \textit{et al.} \cite{Pan2019DiseaseImageSG}  \cite{Pan2021DiseaseimagespecificLF} & FGAN  &  \textcolor{gray}{F} & $L_{gan}$ &  ADNI & MAE, SSIM, PSNR, AUC, ACC, SPE, SEN, F1S, MCC & $\usym{1F5F8}$ &  C \\
        Shin \textit{et al.} \cite{Shin2020GANDALFGA} & GANDALF & \textcolor{gray}{F} & $L_{cgan}$, $L_{p2p}$, $L_{cls}$ & ADNI & ACC, P, R & $\usym{1F5F8}$ & C \\       
        Wei \textit{et al.} \cite{Wei2018LearningMC} & Sketcher-Refiner GAN & \textcolor{gray}{F} &  $L_{cgan}$, $L_{rec1}$ & D13 & DVR &$\usym{1F5F8}$  &  \\    
        Zuo \textit{et al.} \cite{Zuo2021DMCFusionDM} & DMC-Fusion & \textcolor{gray}{F} & $L_{rec1}$, $L_{ssim}$ & D15 & MI, Q, FMI, PSNR, SSIM & $\usym{1F5F8}$  &  \\
        Kao \textit{et al. }\cite{kao2021demystifying} & ESIT &  \textcolor{red}{U} & $L_{rec1}$ & ANDI & MAE, PSNR, SSIM & $\usym{1F5F8}$  & E \\
        Zhang \textit{et al. }\cite{zhang2022bpgan} & BPGAN &  \textcolor{gray}{F} & $L_{gan}$, $L_{rec1}$, $L_{kl}$ & ANDI & MAE, PSNR, SSIM & $\usym{1F5F8}$  & Diag \\
    \hline
    \multicolumn{8}{c}{\textit{CT $\rightarrow$ PET}} \\\hline
           Klser \textit{et al.} \cite{Klser2021ImitationLF} & pPET & \textcolor{gray}{F} &  $L_{rec2}$ & D8 & MAE & $\usym{1F5F8}$ &  \\
            \hline
    
    \multicolumn{8}{c}{\textit{US $\rightarrow$ MRI}} \\\hline
            Jiao \textit{et al.} \cite{Jiao2020SelfSupervisedUT} & \href{https://bitbucket.org/JianboJiao/ssus2mri/src/master}{GAN}& \textcolor{red}{U} & $L_{gan}$, $L_{p2p}$ & D4 & MOS, DS &  $\usym{2717}$ &   \\ 
            \hline
            
    \multicolumn{8}{c}{\textit{PET $\rightarrow$ PET}} \\\hline
           Wang \textit{et al.} \cite{Wang20193DAL} & LA-GANs & \textcolor{gray}{F}  & $L_{gan}$, $L_{rec1}$ & D11, D12 & PSNR, SSIM & $\usym{1F5F8}$ &  SR  \\  
           Zhou \textit{et al.} \cite{Zhou2021SynthesizingMP}  & \href{https://github.com/bbbbbbzhou/UCAN}{UCAN}& \textcolor{gray}{F} &  $L_{gan}$, $L_{iden}$, $L_{cls}$, $L_{cyc}$ & D14 & NMSE, SSIM & $\usym{1F5F8}$ & Diag\\
    \hline
    \end{tabular}
}
\label{tab:summary-methods-2}
\end{table}

\section{Methods}\label{sec:methods}
Table~\ref{tab:summary-methods-1} and Table~\ref{tab:summary-methods-2} provide a taxonomy for the work covered in this survey. Specifically, the second column in the tables is the proposed architecture. The third column denotes the level of supervision, i.e., fully supervised (F), weakly supervised (W), or unsupervised (U) manner. The fourth through sixth columns indicate the loss function, the metrics employed, and whether the input neuroimage images are aligned or misaligned, respectively. The last column refers to the downstream task (DST), which contains super-resolution (SR), segmentation (S), classification (C), detection (D), inpainting (I), registration (R), explainability (E), and diagnosis (Diag). We will discuss each of them in the following sections.

\subsection{Learning Paradigms} \label{sec:learn_paradigm}

\subsubsection{Fully supervised methods} \label{sec:lp_sup}

\revised{Before 2018, most synthesis algorithms adopted the CSC filter~\cite{Zeiler2010DeconvolutionalN} for dictionary learning-based methods. Previously, dictionary learning~\cite{mairal2014sparse} was one of popular models for inverse problems in image de-noising~\cite{giryes2014sparsity,fu2015adaptive}, image super-resolution~\cite{yang2010image,yang2012coupled} and image synthesis~\cite{wang2012semi,huang2013coupled}. For instance, Elad~\textit{et al.}~\cite{elad2012sparse} provide a concise overview of sparse and redundant representation modeling and identify ten important future research directions for sparse coding. Rubinstein~\textit{et al.}~\cite{rubinstein2010dictionaries} explicate the dictionary acquisition process through mathematical models. Gao~\textit{et al.}~\cite{gao2012laplacian} develop a hypergraph Laplacian matrix to retain the local information of the training samples, thereby enhancing the discriminative capacity of the learned dictionary. Li~\textit{et al.}~\cite{li2015locality} employ the locality information to improve the representative ability of the shared dictionary. Chen~\textit{et al.}~\cite{cheng2013sparse} provide an overview of algorithms on sparse representation learning algorithms focusing on visual recognition. However, the major drawback of CSC is that it requires a huge amount of paired data for training. The work in~\cite{Huang2017DOTEDC, Huang2017SimultaneousSA} employs a dual filter training strategy and hetero-domain image alignment to significantly reduce the requirement of a huge amount of paired data.}
When pix2pix was released in 2017, an alternative generative model, GAN, became mainstream for cross-modality neuroimage synthesis~\cite{Wang2018LocalityAM, Siddiquee2019LearningFP}. Zhou~\textit{et al.}~\cite{Zhou2020HiNetHN} pay more attention to the layer-wised fusion strategy from multiple input modalities and design a Mixed Fusion Block (MFB) to combine the latent representation from each source modality. Kwon~\textit{et al.}~\cite{Kwon2019GenerationO3} apply the alpha-GAN to generate a 3D brain MRI from a random vector. Huang~\textit{et al.}~\cite{Huang2019CoCaGANCC} project multi-modality brain MRI data into one common feature space and utilize the modality invariant information represented in the common feature space to generate the missing target domain image space. After that, they apply gradient-weighted class activate mapping (GradCAM)~\cite{Selvaraju2019GradCAMVE} to interpret why the synthesized neuroimages could be utilized for potential clinical usage. Yurt~\textit{et al.}~\cite{Yurt2021mustGANMG} utilize multi-modality neuroimaging data and fuse their features to generate the target domain data. Jog~\textit{et al.}~\cite{Jog2017RandomFR} adopt a multi-scale feature extraction scheme and feed the features to three random forest trees to predict the corresponding area of the target modality. CollaGAN is proposed by~\cite{Lee2019CollaGANCG}, which utilizes the invariant embedding features from multi-modality data and fuses their information to synthesize the target modality data. However, supervised cross-modality neuroimage synthesis algorithms are challenging for a product launch because they need many neuroimage data pairs for training, which is tough to get at medical institutions owing to patients' privacy concerns.

\subsubsection{Weakly supervised methods}\label{sec:lp_semi}

\begin{figure}[t]
    \centering
    \includegraphics[width=0.75\linewidth]{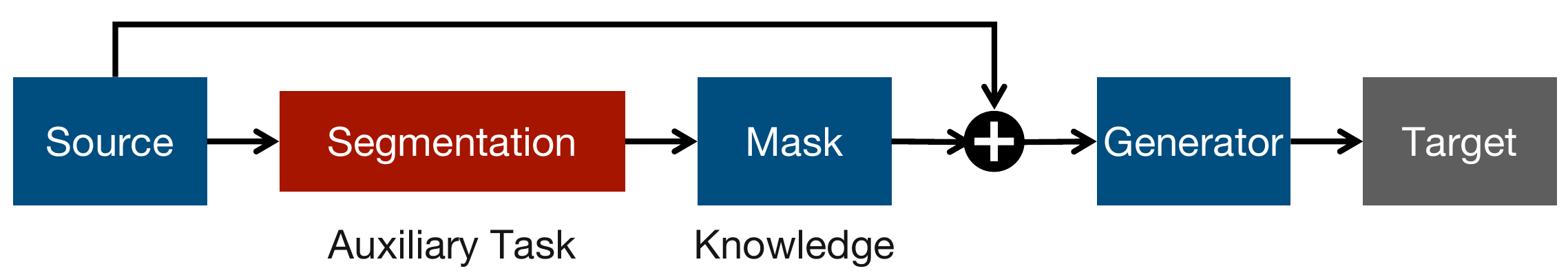}
    \caption{Weakly supervised architecture with an auxiliary task. }
    \label{fig:weak_arch_1}
\end{figure}
\begin{figure}[thb]
    \centering
    \includegraphics[width=0.75\linewidth]{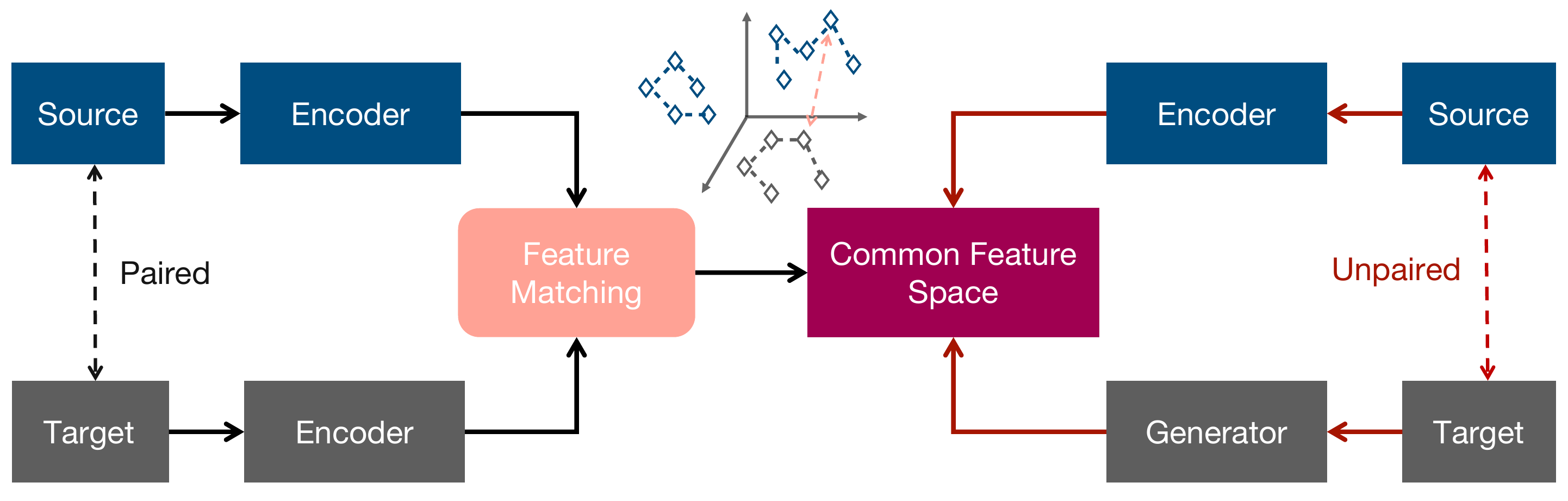}
    \caption{Weakly supervised architecture via projecting into the common feature space.}
    \label{fig:weak_arch_2}
\end{figure}

\revised{We categorize weakly supervised cross-modality neuroimage synthesis algorithms into two streams.} As illustrated in Fig.~\ref{fig:weak_arch_1}, the first stream is to utilize a cross-modality segmentation mask to extract the knowledge from the lesion region~\cite{Shen2021MultiDomainIC}. Then, the knowledge from the segmentation mask is distilled into the generator. In other words, the segmentation network was treated as a teacher to guide the generator using unpaired training data. This method is a potential solution to the problem described in Fig.~\ref{fig:existing_dismerit}. \revised{Since this method transfers the knowledge from the lesion region to the generator}, it is possible to make the generator pay more attention to the lesion region and synthesize a high-fidelity neuroimage. \zz{The second stream~\cite{Huang2018CrossModalityIS,Huang2019SimultaneousSA} aims to make use of a few paired neuroimaging datasets at first and extract the feature from each modality~\cite{zhang2018binary}. BMVC proposed by Zhang~\textit{et al.}~\cite{zhang2018binary} is the first comprehensive work to encode the multi-view image descriptors into a common feature space and clusters via a binary matrix factorization model}. Inspired by this, Huang~\textit{et al.}~\cite{Huang2018CrossModalityIS} project each modality's feature points into the common feature space and adopt maximum mean discrepancy (MMD) to calculate the divergence of paired feature points in the reproducing common feature space. Ultimately, the new feature from the incoming unpaired source modality neuroimage data employs MMD to seek the intrinsic paired feature points in the common feature space. The new target modality neuroimage is generated by feeding the pair feature points from the common feature space into the generator. The flowchart of the second stream method is given in Fig.~\ref{fig:weak_arch_2}. 

\subsubsection{Unsupervised methods}\label{sec:lp_unsup}

\begin{figure}[th]
    \centering
    \includegraphics[width=0.75\linewidth]{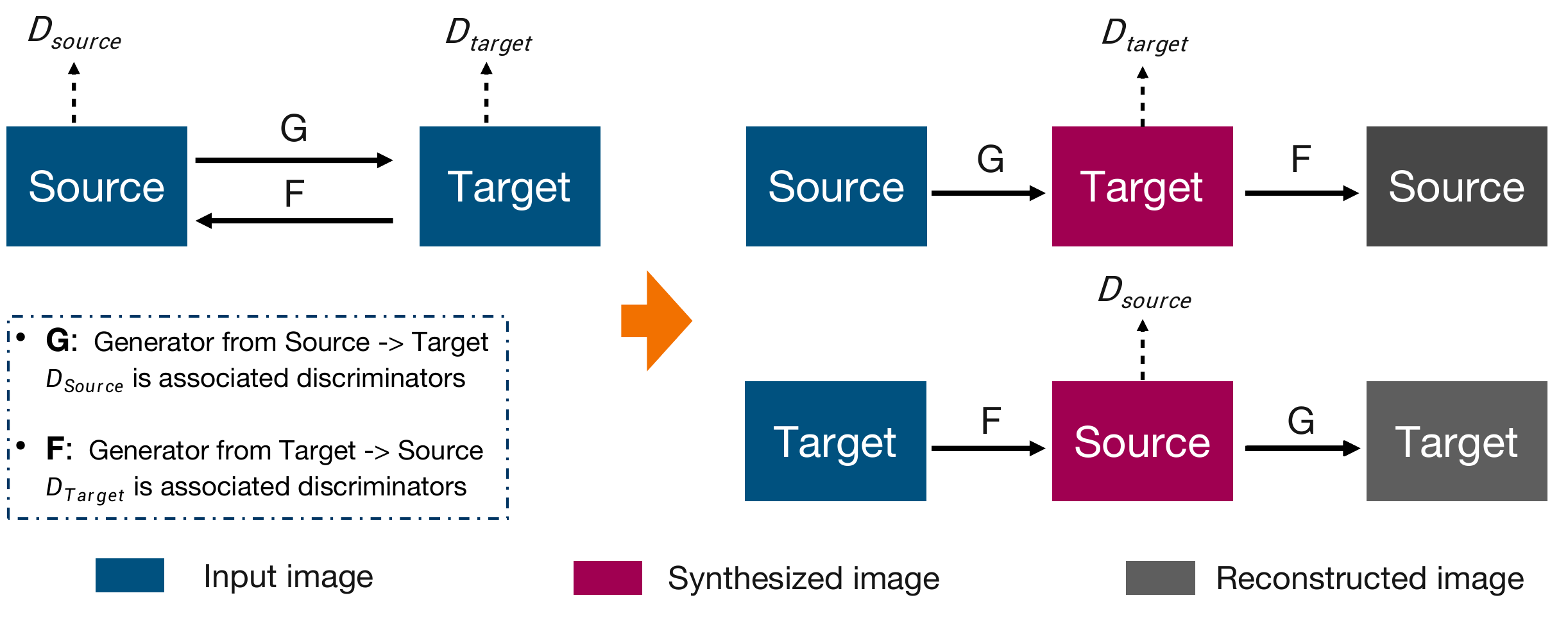}
    \caption{CycleGAN~\cite{Zhu2017UnpairedIT} architecture.}
    \label{fig:cyclegan}
\end{figure}

The mainstream for unsupervised cross-modality neuroimage synthesis is CycleGAN~\cite{Zhu2017UnpairedIT}. In Fig.~\ref{fig:cyclegan}, CycleGAN is separated into two loops. The first loop is source domain$\rightarrow$ synthesized target domain$\rightarrow$ reconstructed source domain. The second loop is target domain$\rightarrow$ synthesized source domain$\rightarrow$ reconstructed target domain. CycleGAN~\cite{Zhu2017UnpairedIT} utilizes two loops to construct a cycle loss function, which is described below: 
\begin{equation}\label{eq:cycle}
    \mathcal{L}_{cycle} = \mathbb{E}_{\mathbf{X} \sim p_{r}(\mathbb{X})}\left \| \mathbf{X} - F(G(\mathbf{X})) \right \|_{1} +  \mathbb{E}_{\mathbf{Y} \sim p_{r} (\mathbb{Y})} \left \| \mathbf{Y} - G(F(\mathbf{Y})) \right \|_{1},
\end{equation}
where $X$ denotes the source domain and $Y$ denotes the target domain; $G$ and $F$ are the generators for task $X \rightarrow Y$ and task $Y \rightarrow X$. The GAN loss function in CycleGAN~\cite{Zhu2017UnpairedIT} is described as below:
\begin{equation}\label{eq:adv}
    \begin{split}
        \mathcal{L}_{adv} &= \mathbb{E}_{\mathbf{Y} \sim p_{r}(\mathbf{Y})}\left [ log\:D_{G}(\mathbf{Y}) \right ] + \mathbb{E}_{\mathbf{X} \sim p_{r}(\mathbf{X})} \left [ log\:(1 - D_{G}(G(\mathbf{X}))) \right ] \\ 
        & + \mathbb{E}_{\mathbf{X} \sim p_{r}(\mathbf{X})}\left [ log\:D_{F}(\mathbf{X}) \right ] + \mathbb{E}_{\mathbf{Y} \sim p_{r}(\mathbf{Y})}\left [ log\:(1 - D_{F}(F(\mathbf{Y}))) \right ].\\
    \end{split}
\end{equation}
Most variants of unsupervised cross-modality synthesis are built on the basis of the cycle loss in Equation~\ref{eq:cycle} and the adversarial loss in Equation~\ref{eq:adv}. Among them, the main difference usually lies in the extraction and alignment of the features~\cite{Huang2020MCMTGANMC,Jiao2020SelfSupervisedUT,Zeng2019HybridGA}. The work presented by Huang~\textit{et al.}~\cite{Huang2020MCMTGANMC} is one of the classical models.

\begin{figure}
    \centering
    \includegraphics[width=0.75\linewidth]{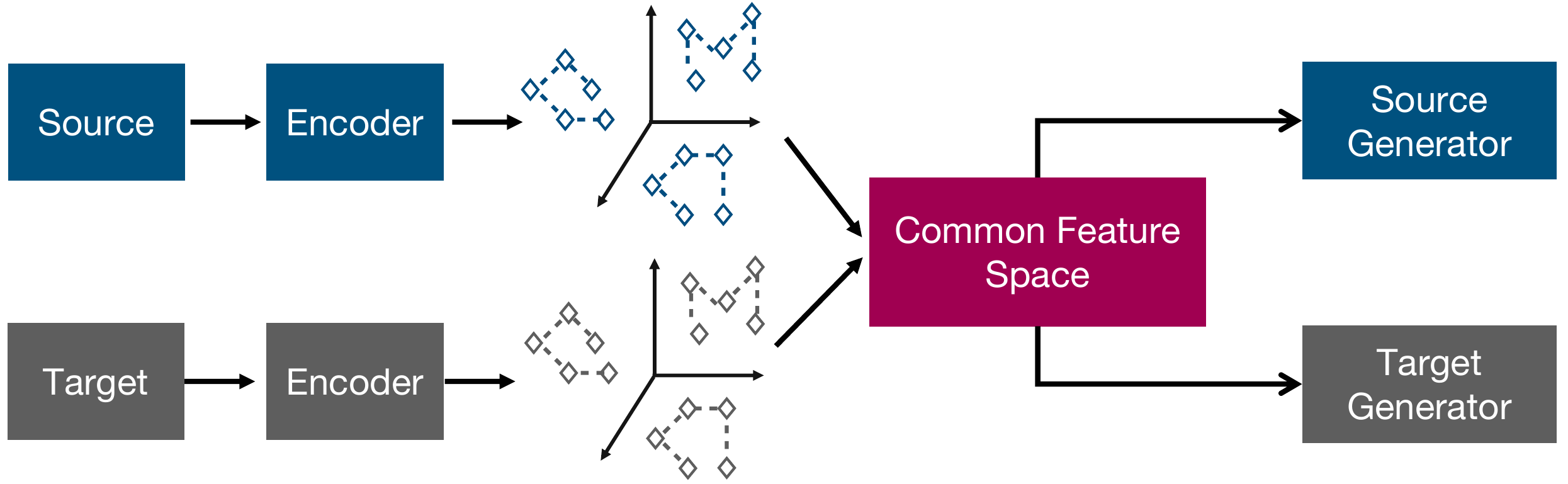}
    \caption{ Unsupervised learning architecture~\cite{Huang2020MCMTGANMC} for cross-modality neuroimage synthesis.  }
    \label{fig:mcmtgan}
\end{figure}

    In Fig.~\ref{fig:mcmtgan}, the unpaired source modality and the target modality neuroimages are fed into the modality-specific feature encoder. \zz{Inspired by~\cite{qin2022joint}, which is the promising work to seek a shared semantic representation but fully preserve the discriminative modal-specific features,} the authors in~\cite{Huang2020MCMTGANMC} utilize extended multi-kernel maximum distance to calculate the distance of features from different modalities. If the distance of features is smaller than a particular threshold, MCMTGAN~\cite{Huang2020MCMTGANMC} aligns them as paired feature points and constructs a common feature space, \zz{which adopts the idea from Zhang~\textit{et al.}~\cite{zhang2022modality}}. In the inference phase, the new unpaired source modality is fed into the modality-specific feature encoder to obtain the reference feature point. Then MCMTGAN \revised{seeks} the paired feature points of the common feature space. The seek feature paired points are fed into the modality-specific generators to synthesize cross-modality neuroimages. Yu~\textit{et al.}~\cite{Yu2021MouseGANGM} present a similar work with the method shown in~\cite{Huang2020MCMTGANMC}. However, the authors pay more attention to the mouse brain dataset. Chen~\textit{et al.}~\cite{Chen2020UnsupervisedBC} propose a more concise idea, i.e., the source modality and the target modality share their feature encoder. Jiao~\textit{et al.}~\cite{Jiao2020SelfSupervisedUT} also extract and map features into the common space using different modalities. Moreover, the authors in~\cite{Jiao2020SelfSupervisedUT} design a new cross-modal attention module for fusion and propagation. Zeng~\textit{et al.}~\cite{Zeng2019HybridGA} use two models; one of which is the 3D generator network, and the other is the 2D discriminator. The authors utilize the result from the 2D discriminator treated as a weak label to supervise the 3D generator, such that the generator's output can be closer to the output of CT. Yang~\textit{et al.}~\cite{Yang2021AUH} design a unified generator for MRI synthesis. The method is also similar to~\cite{Huang2020MCMTGANMC}, which depends mainly on the common feature space. Yang~\textit{et al.}~\cite{Yang2020UnsupervisedMS} also borrow the concept of the common feature space and design a module to make the features from various modalities closer. Tomar~\textit{et al.}~\cite{Tomar2021SelfAttentiveSA} develop a learnable self-attentive spatial normalization with a GAN, which can significantly improve the generator's performance. He~\textit{et al.}~\cite{He2021AutoencoderBS} treat the synthesis problem as a domain generalization problem. The generator's performance on the unseen target modality cannot be guaranteed due to domain-shift problems. 

\subsection{Modality Synthesis Mode}\label{sec:range}

\revised{\subsubsection{MRI contrast synthesis}\label{sec:mri_constrast} The upper part of Table~\ref{tab:summary-methods-1} shows the review for the synthesis across MRI contrasts, like T1-weighted, T2-weighted, and FLAIR. The synthesis across MRI contrasts is a widely discussed subject within the medical neuroimaging community. In the beginning, Bône~\textit{et al.}~\cite{Bne2021ContrastEnhancedBM} and Dar~\textit{et al.}~\cite{Dar2019ImageSI} adopted fully supervised training paradigms for the synthesis across MRI contrasts. Both of them employ autoencoder~\cite{vincent2008extracting} as their neural network architecture. However, a significant limitation of these methods is their reliance on a substantial quantity of paired neuroimaging data for training, which poses a considerable challenge in acquisition within hospitals or medical centers. In 2018, Huang~\textit{et al.}~\cite{Huang2018CrossModalityIS} proposed a semi-supervised method for cross-modality neuroimage synthesis, which fully leverages a large number of unpaired neuroimages to construct a feature space and subsequently extracts the features from a limited number of paired neuroimages to regulate the manifold of feature space. The year 2020 witnessed the emergence of unsupervised cross-modality neuroimage synthesis~\cite{Yang2021AUH} in the academia, as a considerable number of researchers began utilizing GAN for cross-modality neuroimage synthesis~\cite{Sun2020AnAL,Tomar2021SelfAttentiveSA}. For instance, the aim of ANT-GAN~\cite{Sun2020AnAL} is to produce a visually normal neuroimage based on its abnormal counterpart without needing paired neuroimages. Tomar~\textit{et al.}~\cite{Tomar2021SelfAttentiveSA} employ cross-modality neuroimage synthesis to address domain adaptation issues. Furthermore, Yang~\textit{et al.}~\cite{Yang2021AUH} try to propose a unified GAN model for the synthesis of all modalities, including T1-weighted, T2-weighted and FLAIR. Specifically, the authors of~\cite{Yang2021AUH} propose a pair of hyper-encoder and hyper-decoder to facilitate the mapping process from the source contrast to a shared feature space, followed by a subsequent mapping to the target contrast image. }

\subsubsection{MRI To CT}\label{sec:range_mri_ct}
\begin{figure}[t]
    \centering
    \includegraphics[width=0.85\linewidth]{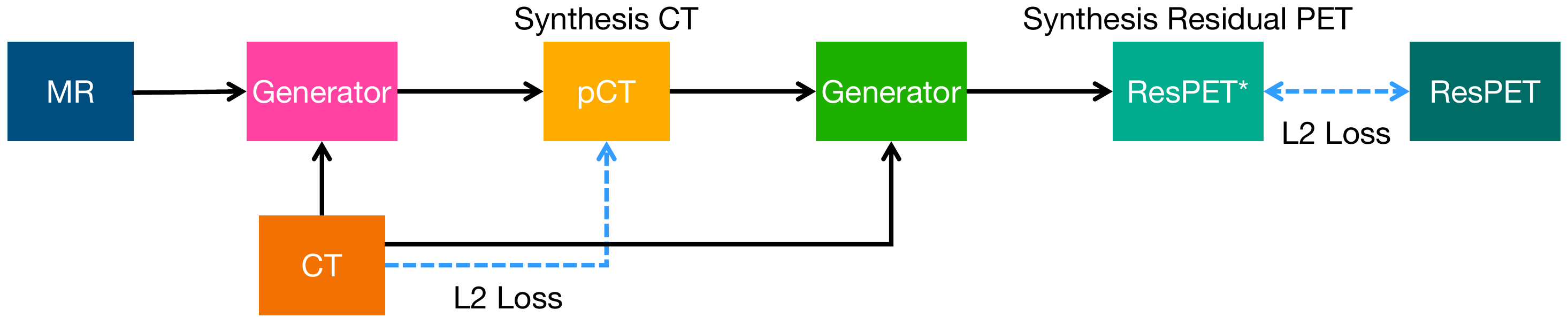}
    \caption{Architecture proposed in~\cite{Klser2021ImitationLF} for MRI to CT and PET synthesis. }
    \label{fig:mri_ct_arch}
\end{figure}

Computed tomography (CT) is of great importance for different clinical applications, such as PET attenuation correction and radiotherapy treatment planning. However, the patients need to be exposed to radiation during CT acquisition, which may cause side effects. Nevertheless, MRI is much safer than CT. Therefore, there is a clear need to synthesize CT~\cite{Nie2017MedicalIS,Huo2019SynSegNetSS,Zeng2019HybridGA} from MRI. In Fig.~\ref{fig:synthesis_range}(a), the first column denotes MRI, and the second column denotes its paired CT. As shown in Fig~\ref{fig:mri_ct_arch}, one of the classical architectures proposed by~\cite{Klser2021ImitationLF} constructs two networks. The role of the first network is to generate CT (pseudo-CT) from MRI. The role of the second network is to generate PET from pseudo-CT. The total training process can be divided into two parts. The first part makes pseudo-CT more consistent with the real CT, and the second part aims to make the generated PET more consistent with the real PET. In the first part of the training stage, the authors feed the paired MR and CT into the first generator. The synthesized CT and the real CT construct an L2 loss function to optimize the parameters of the first generator. In the second part of the training stage, the authors feed the synthesized CT and the real CT into the second generator. The synthesized residual PET and the ground truth residual PET formulate an L2 loss function to optimize the parameters of the second generator. 


\subsubsection{MRI To PET} \label{sec:range_mri_pet}


\begin{figure}
    \centering
    \includegraphics[width=0.85\linewidth]{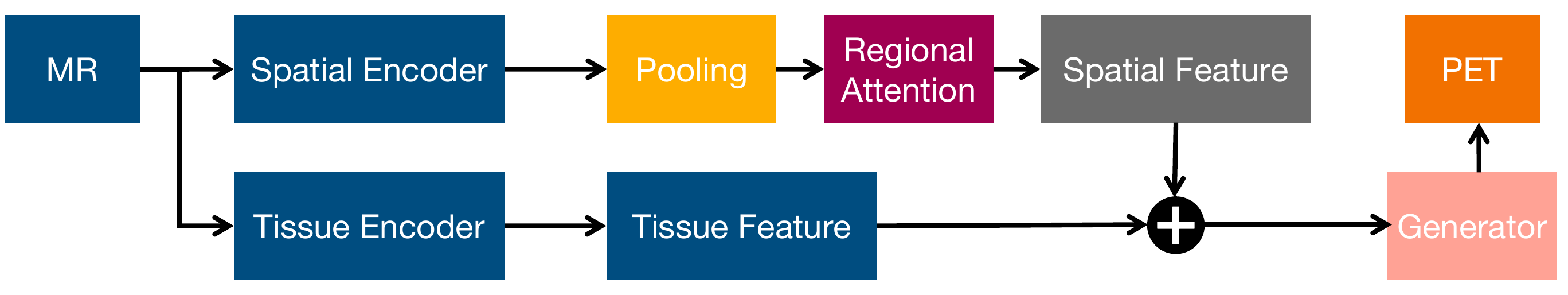}
    \caption{Architecture proposed in~\cite{kao2021demystifying} for MRI to PET synthesis. }
    \label{fig:mri_pet_arch}
\end{figure}

Positron emission tomography (PET) is an essential measure of myelin content changes in vivo in multiple sclerosis. However, PET imaging is costly and invasive due to the injection of a radioactive tracer. In contrast, MRI is much safer since it is not invasive. Therefore, it significantly motivates the researchers to synthesize PET from MRI~\cite{Wei2018LearningMC}. In addition, PET is also regarded as the gold standard for diagnosing Alzheimer's disease (AD). As previously mentioned, PET can be prohibitive due to its cost and invasive nature. Shin~\textit{et al.}~\cite{Shin2020GANDALFGA} \revised{propose} a conditional GAN to synthesize PET from MRI, where the auxiliary information is from AD diagnosis. Liu~\textit{et al.}~\cite{Liu2022AssessingCP} employ a GAN to synthesize PET from MRI and then feed the generated PET and real MRI into the segmentation. Hu~\textit{et al.}~\cite{Hu2022BidirectionalMG} employ a bidirectional mapping mechanism to synthesize between MRI to CT. Kao~\textit{et al.}~\cite{kao2021demystifying} lay the explanatory groundwork for the cross-modal medical image translation model by exploring the biological plausibility behind the deep neural network, mainly focusing on the T1-MRI to PET translation, as shown in Fig.~\ref{fig:synthesis_range}(b). The architecture details are shown in Fig.~\ref{fig:mri_pet_arch}. The input MRI is fed into two feature encoders. One is the spatial encoder; the other is the tissue encoder. The spatial and tissue features are merged and sent into the generator to synthesize the target PET.

\subsubsection{PET} \label{sec:range_pet}


\begin{figure}[th]
    \centering
    \includegraphics[width=0.75\linewidth]{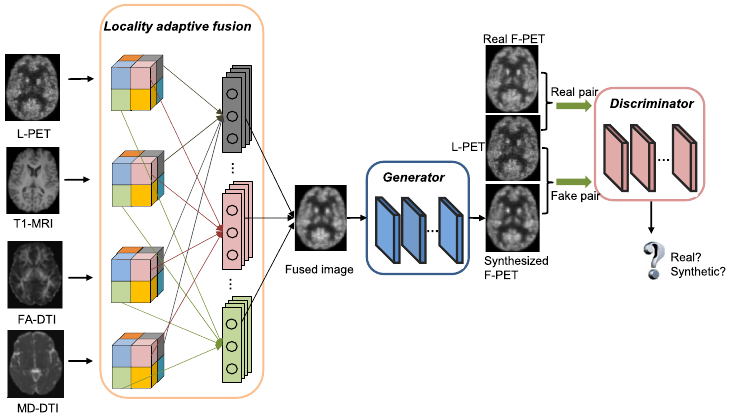}
    \caption{PET synthesis architecture~\cite{Wang20193DAL}.}
    \label{fig:pet_arch}
\end{figure}

Unlike previous methods for one-to-one fixed modality translation, Zhou~\textit{et al.}~\cite{Zhou2021SynthesizingMP} propose a 3D unified Cycle-GAN (UCAN) to synthesize PET, as shown in Fig.~\ref{fig:synthesis_range}(c). Another method is proposed by Wang~\textit{et al.}~\cite{Wang20193DAL}, which is shown in Fig.~\ref{fig:pet_arch}. They propose a 3D auto-encoder to capture various PET modality features into one common space and then utilize the common feature space for synthesizing arbitrary PET modalities. Specifically, the input neuroimage contains PET, MRI, and DTI modalities. The shared feature encoder extracts the feature from modality-specific neuroimage. Then, the feature maps from different modalities are aligned to generate the fused images via the proposed locality adaptive fusion block. The fused images are fed into the generator to synthesize paired neuroimages of different modalities. The role of the discriminator is to classify whether these synthesized paired neuroimages are real or fake. 

\subsubsection{Ultrasound to MRI} \label{sec:range_us_mri}


\begin{figure}[th]
    \centering
    \includegraphics[width=0.75\linewidth]{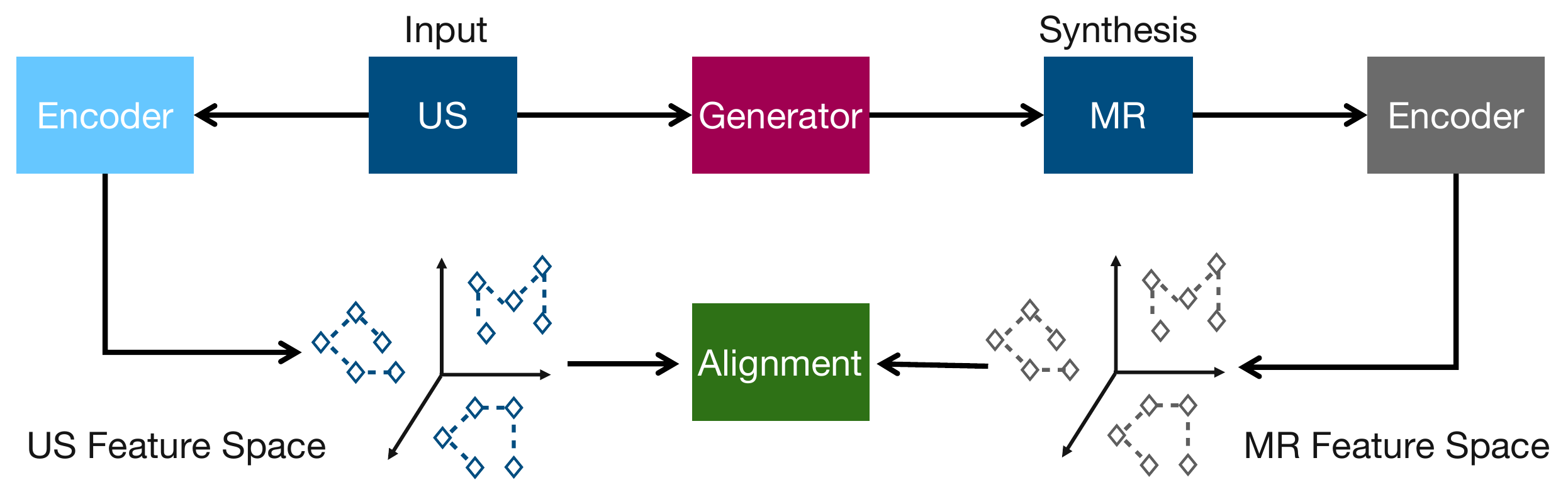}
    \caption{Architecture purposed in~\cite{Jiao2020SelfSupervisedUT} for ultrasound (US) to MRI synthesis.}
    \label{fig:us_pet_arch1}
\end{figure}

Ultrasound (US) is the most common method to detect abnormalities in the fetal brain and growth restriction. However, the quality of ultrasound is easily affected by acoustic windows and occlusions, which mainly come from the fetal brain skull. MRI is unaffected by this case and is able to provide more complete spatial details for full anatomy. One major drawback is that the paired data for ultrasound and MRI is tough to collect. Jiao~\textit{et al.}~\cite{Jiao2020SelfSupervisedUT} employ self-supervised methods to synthesize MRI from ultrasound images, as shown in Fig.~\ref{fig:synthesis_range}(d). In Fig.~\ref{fig:us_pet_arch1}, the ultrasound neuroimage and the MRI neuroimage are fed into the modality-specific feature encoder. All features are aligned in one common feature space. The appearance and edge features of MRI and ultrasound are used to constrain the reconstruction process. 



\subsection{Downstream Tasks} \label{sec:down_task}

\subsubsection{Segmentation} \label{sec:dt_seg} 

\begin{figure}[th]
    \centering
    \includegraphics[width=0.75\linewidth]{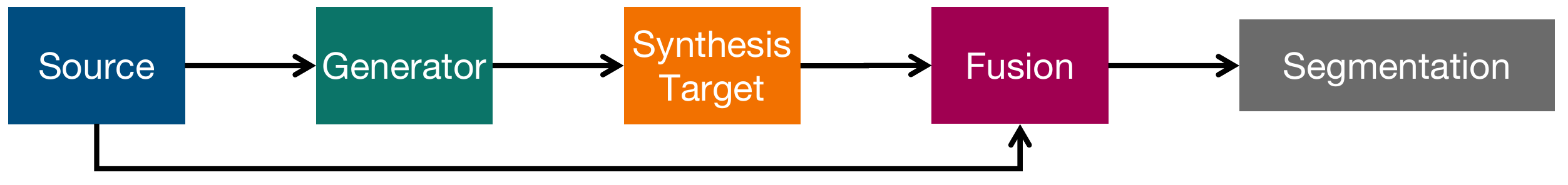}
    \caption{Architecture purposed in~\cite{Shen2021MultiDomainIC}, which uses synthesized images to improve the segmentation accuracy.}
    \label{fig:seg_task1}
    \end{figure}

We summarize how the segmentation task is incorporated into cross-modality neuroimage synthesis. In general, there are two categories of methods. The first category aims to utilize cross-modality synthesized neuroimage to improve the performance of the segmentation task. A typical architecture purposed in~\cite{Shen2021MultiDomainIC} is shown in Fig.~\ref{fig:seg_task1}. The synthesized target and original source modality neuroimage are concatenated and fed into the segmentation network~\cite{Huo2019SynSegNetSS, Yu2021MouseGANGM, Shen2021MultiDomainIC}. Huo~\textit{et al.}~\cite{Huo2019SynSegNetSS} directly use the accuracy of the segmented results to evaluate whether the synthesized data is helpful without evaluating the quality of the synthesized results by peak signal-to-noise ratio (PSNR) and the structure similarity index (SSIM). Yu~\textit{et al.}~\cite{Yu2021MouseGANGM} jointly optimize the synthesis and segmentation tasks using an unsupervised learning method.

\begin{figure}[th]
    \centering
    \includegraphics[width=0.75\linewidth]{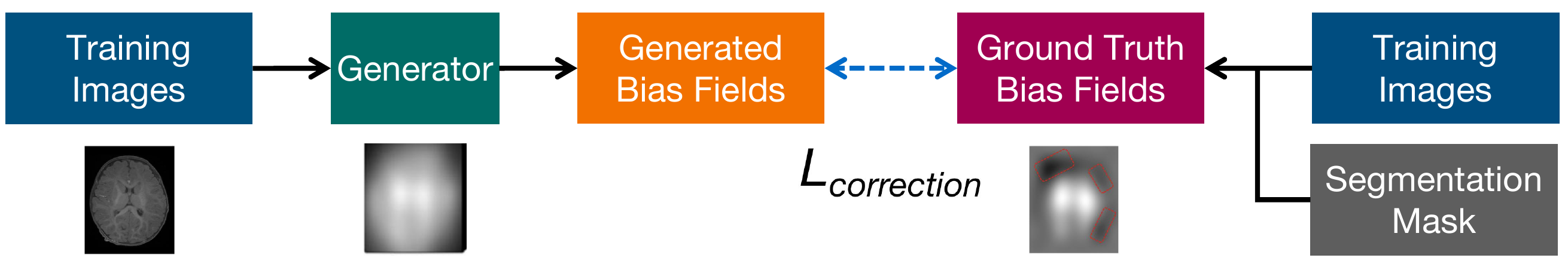}
    \caption{Architecture purposed in~\cite{Chen2021ABCnetAB}, which uses segmentation masks to improve the quality of synthesized images. }
    \label{fig:seg_task2}
    \end{figure}
   
Furthermore, the second category aims to utilize a segmentation mask to boost the performance of cross-modality neuroimage synthesis~\cite{Chen2021ABCnetAB, Guo2021AnatomicAM, Zhou2021AnatomyConstrainedCL}. For instance, the architecture of Chen~\textit{et al.}~\cite{Chen2021ABCnetAB} is shown in Fig.~\ref{fig:seg_task2}. To be specific, the training images and the corresponding segmentation masks generate the ground truth bias fields. Then, the synthesis bias fields and the ground truth bias fields formulate a correction loss to optimize the parameters of the generator. Chen~\textit{et al.}~\cite{Chen2021ABCnetAB} pay more attention to the brain MRI of infants. The authors incorporate the manual annotations of tissue segmentation maps into the synthesis procedure and make the generated data more segmentation-oriented. Finally, the work in ~\cite{Chen2021ABCnetAB} proves that the synthesized maps can significantly improve segmentation accuracy. Guo~\textit{et al.}~\cite{Guo2021AnatomicAM}, Shen~\textit{et al.}~\cite{Shen2021MultiDomainIC} and Zhou~\textit{et al.}~\cite{Zhou2021AnatomyConstrainedCL} leverage the segmentation task to guide the synthesis task.

\begin{figure}[th]
    \centering
    \includegraphics[width=0.75\linewidth]{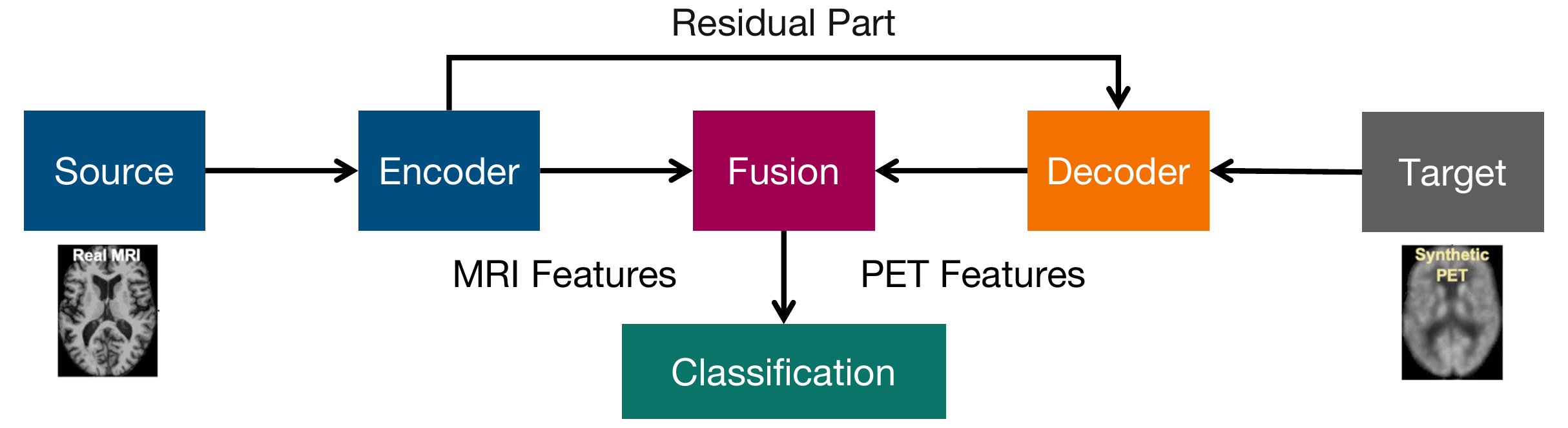}
    \caption{Architecture purposed in~\cite{Liu2022AssessingCP} for cross-modality neuroimage synthesis for the downstream classification task.}
    \label{fig:classification}
\end{figure}

\subsubsection{Classification} \label{sec:dt_class}
In Fig.~\ref{fig:classification}, we summarize how the classification task is incorporated into cross-modality neuroimage synthesis in the method purposed in~\cite{Liu2022AssessingCP}. The feature embedding from the source modality neuroimage and the feature embedding from the synthesis target modality are mixed as the input for the classification task~\cite{Liu2022AssessingCP,Shin2020GANDALFGA,Hu2020BrainMT}. Similar to Fig.~\ref{fig:classification}, the work in~\cite{Shin2020GANDALFGA} incorporates AD's information as an auxiliary task to improve the target modality image synthesis performance. Since the synthesis process is classification-oriented, the synthesized brain image can primarily improve the performance of AD's classification. Hu~\textit{et al.}~\cite{Hu2020BrainMT} design a bidirectional mapping mechanism to preserve the brain structures with high-fidelity details. The work in~\cite{Hu2020BrainMT} verifies that the synthesized neuroimaging data can increase the classification accuracy. 



\begin{figure}[th]
    \centering
    \includegraphics[width=0.75\linewidth]{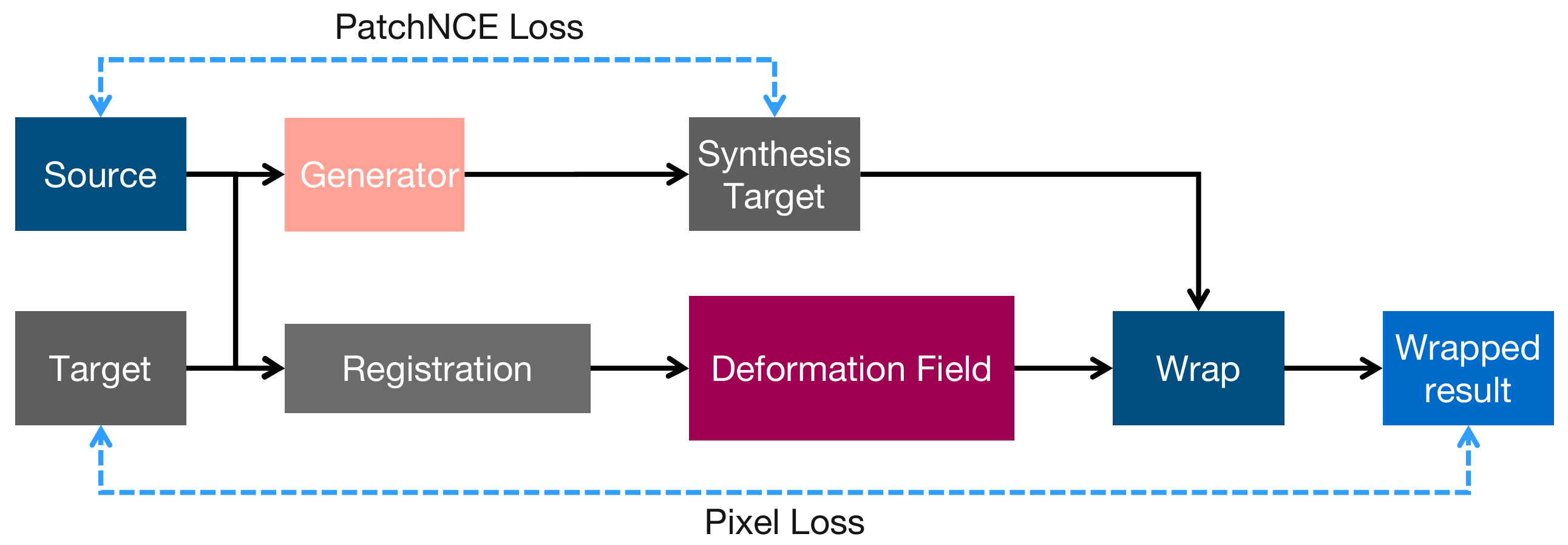}
    \caption{Architecture proposed in~\cite{Chen2022UnsupervisedMM} for cross-modality neuroimage synthesis for the downstream registration task.}
    \label{fig:registration}
\end{figure}

\begin{figure}[t]
    \centering
    \includegraphics[width=0.75\linewidth]{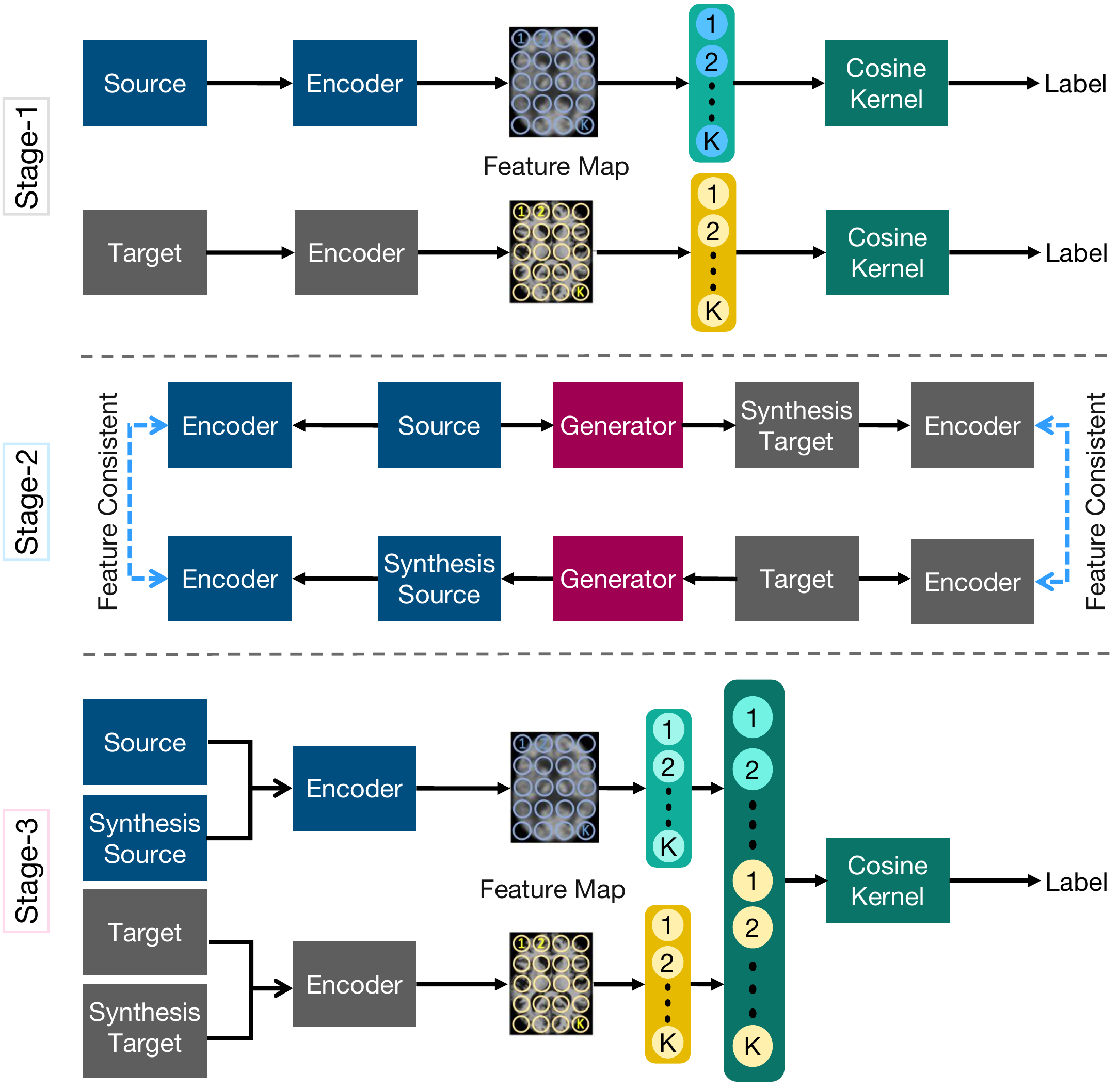}
    \caption{Architecture proposed in~\cite{Pan2021DiseaseimagespecificLF} for cross-modality neuroimage synthesis for the downstream diagnosis task~\cite{Pan2021DiseaseimagespecificLF}.}
    \label{fig:diagnose_stage}
\end{figure}

\subsubsection{Registration}

Multi-modality neuroimage registration~\cite{Maes1997MultimodalityIR} is a traditional topic for the medical imaging community. However, the upcoming trend is utilizing cross-modality neuroimage synthesis for registration~\cite{Chen2022UnsupervisedMM,He2022NonfinitemodalityDA}. One representative example is given in ~\cite{Chen2022UnsupervisedMM} and its architecture detail is shown in Fig.~\ref{fig:registration}. The model consists of two components. One is the registration network, and the other is the cross-modality synthesis network. Both of them are jointly optimized. The source modality neuroimage is warped to align with the target image via the deformation field. Then, the authors propose a novel PatchNCE loss to preserve the shape information in the synthesized neuroimage. Furthermore, Chen~\textit{et al.}~\cite{Chen2022UnsupervisedMM} employ a pixel-wise reconstruction loss function to preserve the appearance of the synthesis neuroimage. In this case, the proposed model can simultaneously optimize the generator and the registration network.

\subsubsection{Diagnosis} \label{sec:dt_diagn}




Pan~\textit{et al.}~\cite{Pan2021DiseaseimagespecificLF} were the first to apply cross-modality neuroimage synthesis to disease diagnosis tasks. The training process is divided into three stages. The architecture details are described in Fig.~\ref{fig:diagnose_stage}. In the first stage, the input source modality neuroimage and its paired target modality neuroimage are fed into the feature encoder to obtain the modality-specific feature map. They then propose a spatial cosine kernel function to decompose the output feature map into two \revised{parts}. One is the disease-relevant part, and the other is the residual normal part, which is irrelevant to the disease. The exploited spatial cosine kernel is to ensure that the classifier can learn the disease-relevant features. In the second stage, the source and target feature encoders are frozen. In other words, only the generator is trained in an end-to-end manner. The second stage aims to encourage the feature map from the synthesized neuroimage to be consistent with the feature map from the input neuroimage. In the third stage, the features from the source modality neuroimage and the target modality neuroimage are concatenated for brain disease identification. However, this work adopts a supervised learning method by inputting paired multi-modality neuroimaging data. It is challenging to apply to other tasks since it is very difficult to collect fully paired data.

\begin{table}[t]
\caption{A brief summary of datasets for cross-modality neuroimage synthesis. Note that \textcolor{blue}{$^{\sharp}$} and \textcolor{red}{$^{\natural}$} indicate that it has labels for the segmentation task and classification task, respectively.}
\renewcommand{\arraystretch}{1.2}
\resizebox{\textwidth}{!}{
    \begin{tabular}{p{2.5cm}p{2cm}p{3cm}p{14cm}}
    \hline
    \textbf{Dataset} & \textbf{Subjects} & \textbf{Modality} &  \textbf{Data Description} \\\hline

    \multicolumn{4}{c}{\textit{Public}} \\\hline
    
    IXI & 578 & MR (T1w, T2w, PDw) &  Healthy subjects, 3 hospitals with different acquisition protocols, volume size: $256 \times 256 \times z $ ($z = 112-136$). \\
    & & & URL: \href{http://brain-development.org/ixi-dataset}{http://brain-development.org/ixi-dataset}. \\

    BRATS15~\cite{menze2014multimodal} & 65\textcolor{blue}{$^{\sharp}$}   & MR (T1, T1c, T2, FLAIR)  & Multi-contrast MR scans from glioma patients. URL: \href{https://www.smir.ch/BRATS/Start2015}{https://www.smir.ch/BRATS/Start2015}. \\
    
    BRATS18~\cite{chartsias2017multimodal} & 285\textcolor{blue}{$^{\sharp}$}   & MR (T1w, T2w, FLAIR) & 210 high-grade glioma (HGG) and 75 lower grade glioma (LGG) MRI with binary masks for the tumor (or lack of tumor). Each 3D MRI contains 155 slices of size $240 \times 240$. URL:  \href{https://www.med.upenn.edu/sbia/brats2018.html}{https://www.med.upenn.edu/sbia/brats2018.html}. \\
    
    BRATS21  &  2000\textcolor{blue}{$^{\sharp}$}   & MR (T1, T2, FLAIR) &   URL: \href{http://www.braintumorsegmentation.org}{http://www.braintumorsegmentation.org}.\\
    
    NAMIC &  20 & MR (T1w, T2w) & 10 normal controls and 10 schizophrenic, volume size: $128\times128\times z$ ($z = 88$). URL: \href{https://www.na-mic.org/wiki/Downloads}{https://www.na-mic.org/wiki/Downloads}. \\
    
    HCP~\cite{van2013wu} & 200 & MR (T1w, T2w) & The Human Connectome Project (HCP), but for healthy young-adult subjects. \\
    & & & URL: \href{https://www.humanconnectome.org}{https://www.humanconnectome.org}. \\
    
    
    BCP~\cite{howell2019unc} & 546 & MR (T1w, T2w) & Real infant MRI scans. \\
    
    ProstateX~\cite{litjens2014computer} & 98 & MR (T2w, ADC, DWI) & Contain 3 modalities: T2w, apparent diffusion coefficient (ADC) and high b-value DWI images. \\
    
    ISLES15~\cite{maier2017isles} & 65\textcolor{blue}{$^{\sharp}$} & MR (T1w, T2w, FLAIR, DWI) &
    Ischemic stroke lesion segmentation (ISLES), a medical image segmentation challenge on MICCAI 2015. URL: \href{http://www.isles-challenge.org/ISLES2015}{http://www.isles-challenge.org/ISLES2015}. \\
    
    MIDAS~\cite{Dar2019ImageSI}&  66 & MR (T1w, T2w) & 48 subjects for training, 5 for validation and 13 for testing.\\
    
    ANDI~\cite{jack2008alzheimer} & 680\textcolor{red}{$^{\natural}$}  & MR (T1w), PET  & Alzheimer's Disease Neuroimaging Initiative (ADNI).\\
    & & & URL: \href{https://adni.loni.usc.edu}{https://adni.loni.usc.edu}. \\
    
    MMRR~\cite{landman2011multi}& 21 & MR (T2w) & Multimodal Reproducibility Resource (MMRR).\\ 
    
    ATLAS~\cite{liew2018large}& 220 & MR (T1w) & Anatomical tracings of lesions after stroke (ATLAS), for stroke MRI generation. \\
    & & & URL: \href{http://fcon_1000.projects.nitrc.org/indi/retro/atlas.html}{http://fcon\_1000.projects.nitrc.org/indi/retro/atlas.html}. \\
    
    CLAS  & 76 & MR (T1w) & Chinese Longitudinal Aging Study (CLAS). \\
    
    AIBL~\cite{ellis2009australian}& 235 & MR (T1w), PET & Australian imaging, biomarkers and lifestyle (AIBL) with paired T1w MRI and  Flute/PIB-PET scans.\\
    
    MRBrainS13 \cite{mendrik2015mrbrains} & $20^{\sharp}$ & MR (T1, FLAIR) &  URL: \href{https://mrbrains13.isi.uu.nl}{https://mrbrains13.isi.uu.nl}. \\
    
    MRBrainS18 & $7^{\sharp}$ & MR (T1, FLAIR) &  Providing ground truth labels for 10 categories of brain structures. \\ 
    & & & URL: \href{https://mrbrains18.isi.uu.nl}{https://mrbrains18.isi.uu.nl}. \\
    
    iSeg17~\cite{wang2015links}&  23\textcolor{blue}{$^{\sharp}$} & MR (T1, T2) & 661 training images, 163 images for testing. \\
    
    RIRE~\cite{west1997comparison} & 19 & MR (T1, T2) & 
    Including T1 and T2 images collected from 19 subjects. \\

    \hline
    \multicolumn{4}{c}{\textit{Private}} \\\hline

    D1~\cite{Hemsley2020DeepGM} & 105 & MR (T1w, FLAIR), CT & Image size $512 \times 512$. \\
    
    D2~\cite{Wang2018LocalityAM}  & 16 & MR (T1), PET & 8 normal control subjects and 8 mild cognitive impairment subjects, each with a low-dose PET image, a T1-MRI image and a full-dose PET image. Each aligned image has the resolution of $2.09\times2.09\times2.03 $ mm$^3$ and the volume size of $128\times128\times128$.\\
    
    D3~\cite{Huo2019SynSegNetSS}  & 60MRI, 19CT & MR (T2w), CT &In total, 3262 MRI slices and 1874 CT slices were used in the experiments.\\
    
    D4~\cite{Jiao2020SelfSupervisedUT} & 107US, 2MRI & MR, US & Around 36,000 2D US slices and 600 MRI slices were extracted accordingly.\\
    
    D5~\cite{Guo2021AnatomicAM}  & 100 & MR (T1w, T2w, FLAIR) & Volume size:$256 \times 255 \times 15$, 3 labels: edema; cavity; and tumor.\\
   
    D6~\cite{Yang2020UnsupervisedMS} & 45 & MR, CT & Training set of 28 subjects, validation set of 2 subjects, and test set of 15 subjects.\\
    
    D7~\cite{Zeng2019HybridGA}  & 50 & MR, CT & 2 modalities are aligned with a rigid registration, volume size: $256 \times 288 \times 112$. \\
    
    D8~\cite{Klser2021ImitationLF} & 20MRI, 18PET & MR (T1w, T2w), PET & 20 pairs of brain MR, CT and 18F-FDG PET images.\\
    
    D9~\cite{Lee2019CollaGANCG}  & 10 & MR (T1F, T2w, T2F) & Total 280 axial brain images were scanned and additional T2 FLAIR sequences from 10 subjects. There are four types of MR contrast: T1-FLAIR, T2w, T2-FLAIR, and T2-FLAIR$^*$.\\
    
    D10~\cite{Li2019DiamondGANUM}  & 65 &  MR (T1, T2, FLAIR, DIR) & 65 scans of patients with MS lesions from a local hospital.\\ 
    
    D11~\cite{Wang20193DAL}  & 20&  MR (T1), PET & 20 simulated subjects were generated from the BrainWeb database of twenty normal brains. \\
    
    D12~\cite{Wang20193DAL}  & 16 &  MR (T1), PET &  8 normal control subjects and 8 mild cognitive impairment subjects, each with L-PET, F-PET, T1-MRI, FA-DTI and MD-DTI.\\
    
    D13~\cite{Wei2018LearningMC} & 18 &  MR, PET & 18 patients (12 women, age 31.4 $\pm$ 5.6) and 10 age- and gender-matched healthy volunteers (8 women, age 29.4 $\pm$ 6.3).\\ 

    D14~\cite{Zhou2021SynthesizingMP}  & 35 & PET  & 24 patients diagnosed with AD and 11 patients as healthy control.\\
    
    D15~\cite{Zuo2021DMCFusionDM}  & 74 pairs & NR (T1, T2), CT, PET & 14 pairs of CT/MRI images, 20 pairs of MR-T1/MR-T2, 20 pairs of MR/SPECT, 20 pairs of MR/PET images.\\
    
    D16~\cite{al2022unpaired}  & 20 & MR, CT & Unpaired brain MR and CT volumes, including 179 two-dimensional (2D) axial MR and CT images.\\
     
    \hline
    \end{tabular}
}
\label{tab:summary-dataset}
\end{table}

\section{Datasets, Losses and Metrics}\label{sec:datast_loss_metric}

\subsection{Datasets} \label{sec:dataset}
Table~\ref{tab:summary-dataset} presents a summary of the public datasets. It can be easily observed that most public datasets can only be used synthesis for different MRI contrasts (namely, MRI-to-MRI synthesis). There are no datasets for MRI to PET, MRI to CT, or CT to PET synthesis. Most deep-learning models require considerable data to avoid model over-fitting. Hence, it is an urgent issue that there is not enough data to support the research for cross-modality synthesis except for MRI-to-MRI synthesis. In addition, according to Fig.~\ref{tab:summary-dataset}, the publicly available MRI datasets are quite small. The largest dataset, BraTS~\cite{Chartsias2018MultimodalMS}, only contains 3000 subjects. Most MRI-to-MRI datasets contain no more than 500 subjects. Moreover, we can find that a lot of public MRI-to-MRI datasets have no ground truths for downstream tasks, like segmentation and classification. Since cross-modality synthesis aims to serve the downstream task, it is not easy to evaluate the real value of the synthesized neuroimage without the label for the downstream task. Most evaluation metrics, like PSNR and SSIM, are proposed to measure the quality of natural images, which cannot reflect the quality of cross-modality neuroimages. Thus, the performance of downstream tasks is a more robust metric for cross-modality neuroimage synthesis~\cite{Huang2020MCMTGANMC}.




\begin{table*}[t]
\caption{A brief summary of different losses for cross-modality neuroimage synthesis}
\renewcommand{\arraystretch}{1.2}
\resizebox{\textwidth}{!}{
    \begin{tabular}{lp{6cm}p{14cm}}
    \hline
    \textbf{Abbr.} & \textbf{Function} & \textbf{Description} \\\hline

    $L_{gan}$ & $  E_I[\log D(I)] + E_I [ \log (1 - D(G(I)))]$ & $G$ and $D$ are the generator and discriminator, respectively.\\
    $L_{cgan}$ & $  E_{x, y}[\log D(x, y)] + E_{x, z}[\log (1 - D(x, G(x, z)))]$ & Learn the transformation from conditioned sample $x \in X $ and random noise vector $z$ to the desired output sample $y \in Y, G:\{x, z\} \rightarrow \{y\}$. G and D are the generator and discriminator, respectively, and $E_{a, b}[f(x)]$ is the expected value of $f (x)$ over the distributions of $a$, $b$. \\
    $L_{sgan}$ & $  E_{y\sim \mathcal{Y}}[ D_{\mathcal{Y}}(y)^2] + E_{x\sim \mathcal{X}}[ (1 - D_{\mathcal{Y}}(G_{\mathcal{Y}}(x)))^2]$ & $G$ and $D$ are the generator and discriminator, respectively.\\
    $L_{p2p}$ & $ E_{x, y, z}[|y - G(x, z)|_1]$ &  In the context of image-to-image translation, low-frequency information is better captured when an $l_1$ penalty $|y - G(x,z)|_1$ is added to the loss function. \\
    $L_{cont}$ & $E_{x \sim p}[||\phi(G(x))-\phi(x)||_1]$ & This loss is to ensure that the resulting data achieved from generator $G$ retains the same content as the input. $\phi(\cdot)$ represents feature maps, and it adopts an $l_1$ distance to measure the cross-quality content loss. \\
    $L_{prec}$ & $E_{x, y}[||V(y)-V(G(x))||_1]$ & The perceptual loss is to ensure that the resulting data achieved from generator $G$ retains the same content as the output. $V(\cdot)$ represents feature maps, and it adopts an $l_1$ distance to measure the cross-quality content loss. Incorporating a perceptual loss during network training can yield visually more realistic results.\\
    $L_{text}$ & $ \sum_{l}||T_l(\phi(F(\hat{I}^L)))-T_l(\phi(I^H))||^2_2$ & A texture descriptor \cite{ulyanov2016texture}, involving several Gram matrices $T_l$.\\
    $L_{cyc}$ & \makecell[l]{$E_{x\sim A}[||G_B(G_A(x))-x||_1] +$ \\ $ E_{y\sim B}[||G_A(G_B(y))-y||_1]$}  &  Enforce forward-backward transformation consistency.\\
    $L_{seg}$ & $-\sum_i \log(Seg(G_1(x_i)))$ &  The segmentation loss is the weighted cross entropy loss. $Seg(\cdot)$ is the segmentation network.\\
    $L_{seg1}$ & $E_{x,c}[S(G(x,c), gt)]$ &  We concatenate the corresponding modality label $c$ to the synthesized image as the input of the segmentation network $S$. $gt$ is the ground truth of the tumor segmentation map.\\
    
    $L_{iden}$ &  $E_{y\sim B}[||G_Y(y)-y||_1]$ & Identity consistency constraint, which can regularize the generator to preserve the colors and intensities during translation.\\
    
    $L_{rec1}$ &  $E_{x\sim A, y \sim B, z\sim P(z)}||y-G(x, z)||_1$ & The image $y$ is reconstructed by generator $G$ using the input $x$ with the $l_1$ constraint. $P(z)$ is the standard normal distribution and $z\sim P(z)$ is the sampled latent vector. \\
    
    $L_{rec2}$ &  $E_{x\sim A, y \sim B, z\sim P(z)}||y-G(x, z)||_2$ & Similar with $L_{rec}$ but uses $l_2$ constrain. \\
    
    $L_{kl}$ & $\mathbb{E}[D_{KL}(E(y))||N(0, 1)]$ & To ensure the encoded vector has a similar distribution to the sampled latent vector, the KL-divergence constraint is enforced in the encoder network. It means that the difference between the encoded vector and the latent vector should be minimized. $\mathbb{E}$ denotes the expected value, $E$ represents the encoder, $D_{KL}$ denotes KL divergence, and $z$ represents the latent vector sampled from the standard normal space.\\
     
    $L_{cla}$ & \makecell[l]{$L_{CE}(C(E(G(I_i, c_i), c_i), c_j), c_j)+$ \\ $ L_{CE}(C(E(I_i, c_i), c_i))$} & Contrast-classification loss is to force classifier $C$ to predict the contrast of extracted features by encoder $E$. We adversarially train the classifier to make deep features of multiple contrasts extracted by $E$ to be same distributed, i.e., within a common feature space, using a gradient reversal layer in $C$, which flips gradient sign during backpropagation to force extracted deep features unable to be classified by $C$. $L_{CE}$ computes the cross entropy between estimated contrast probability by $C$ and real contrast code.\\
     
    $L_{cls}$ & $-y \log(x) - (1-y)\log(1-p(x))$ & Universal cross-entropy loss. $p(x)$ is the estimated probability of $x$ belonging to the correct class $y$.\\
    
    $L_{am}$ & $E_{p_a(x)}[||(\mathbf{1} - \mathrm{M}_x) \odot (G(x^a)-x^a)||^2_2]$ & For this medical imaging task in which accuracy is a major requirement of the model, the generator $G$ needs to detect and modify the lesion region while keeping other parts unchanged. The $L_{am}$ penalty is meant to enforce this, but to further help in this task we include a global shortcut to require the generator to learn a mapping that isolates and removes the lesion. $\mathrm{M}_x$ is mask of $x$. $\odot$ represents element-wise multiplication and $\mathbf{1}$ is an all-ones matrix of the same size as the input image.\\
    
    $L_{fm}$ & $\sum_{k}\sum_i \frac{1}{N_i}||D^i_k(x, y) - D^i_k(x, G(x))||^2_2$ & Feature matching loss is to stabilize training, which optimizes the generator to match these intermediate representations from the real and the synthesized images in multiple layers of the discriminators. $D^i_k$ denotes the $i$th layer of the discriminator $D_k$, and $N_i$ is the number of elements in the $i$th layer.\\
    
    $L_{sc}$ & $\mathrm{GDL}(s, U(y)) + \mathrm{GDL}(s, U(G(x)) $ & Shape consistency loss is to regularize the generator to follow consistency relations. It adopts a generalized Dice loss (GDL) \cite{sudre2017generalised} to measure the difference between the predicted and real segmentation maps. $U(y)$ and $U(G(x))$ represent the predicted lesion segmentation probability maps by taking $y$ and $G(x)$ as inputs in the segmentation module, respectively. $s$ denotes the ground truth lesion segmentation map.\\
    
    $L_{cm}$ & $c \otimes ||\hat{y} - y||_1 - \lambda \sum_i \sum \log(c^{ij})$ & Confident map loss is to model the data-dependent aleatoric uncertainty. $\hat{y}$ is intermediate synthesis results of synthesized image $y$. $c$ is the confidence map. To avoid a trivial solution (i.e. $c^{ij}=0,\forall i,j$), $\lambda$ is a constant adjusting the weight of this regularization term.\\

    \hline
    \end{tabular}
}
\label{tab:summary-loss-first}
\end{table*}

\begin{table*}[ht]
\caption{A brief summary of different losses for cross-modality neuroimage synthesis (continuation of Table 5)}
\renewcommand{\arraystretch}{1.2}
\resizebox{\textwidth}{!}{
    \begin{tabular}{lp{6cm}p{14cm}}
    \hline
    \textbf{Abbr.} & \textbf{Function} & \textbf{Description} \\\hline
    $L_{dice}$ & $1-\frac{1}{L}\sum^L_{l=1}\frac{\sum_p 2\hat{y}_p y_p(l)}{\hat{y}_p(l)^2 + \sum_p y_p(l)^2}$ & Dice loss \cite{milletari2016v} measures the segmentation accuracy, where $L$ is the total number of classes, $p$ is the spatial position index in the image, $\hat{y}(l)$ is the predicted segmentation probability map for class $l$ from the segmentation network and $y(l)$ is the ground truth segmentation mask for class $l$.\\

    $L_{reg}$ & \makecell[l]{
    $\mathrm{E}_{x\sim \mathcal{X}}\left [\left\| \mathcal{A}_{\mathcal{X}}(x)\mathcal{A}_{\mathcal{X}}(x)^{T}-1\right\|_{F}\right ] + $ \\
    $\mathrm{E}_{y\sim }\left [\left\|\mathcal{A}_{\mathcal{X}}(G_{\mathcal{X}}(y))\mathcal{A}_{\mathcal{X}}(G_{\mathcal{X}}(y))^{T} -1 \right\|  \right ]_{F}$} &  Attention regularization loss term encourages the attention maps to be orthogonal to each other. $\mathcal{A}$ the attention module. $1$ is the identity matrix and $||\cdot||_F$ denotes the Frobenius norm.\\
    
    $L_{liu}$ & $\frac{1}{n}\sum^n_{i=i}\frac{\sigma(p_i)}{\mu(p_i)} + \frac{1}{n}\sum^n_{i=i}\frac{\sigma(q_i)}{\mu(q_i)}$ & To well handle the spatiotemporally-heterogeneous intensity changes of the brain MR images, $L_{liu}$ is designed to encourage the local intensity homogeneity in the corrected image. We assume that, for each brain tissue, i.e., gray matter (GM) and white matter (WM), the intensity values of the corrected image within a local patch are relatively homogeneous. $p_i$ and $q_j$ denote the $i$th and $j$th local patches sampled from the GM and WM of the “corrected” image, respectively. The operators $\sigma(\cdot)$ and $\mu(\cdot)$ are the standard deviation and mean, respectively. $\sigma(\cdot)/\mu(\cdot)$ calculates the coefficient of variation for a local patch.\\
    
    $L_{cor}$ & $\gamma(1-\kappa_{\zeta}(G(v), f))$ &  $L_{cor}$ can more sensitively suppress the negative influence from the outliers or impulsive noises, and thus help stabilize the training procedure and consequently improve the quality of the bias fields. $\kappa_{\zeta}(G(v), f)=\exp(-\frac{(G(v)-f)^2}{2\zeta^2})$ is the Gaussian kernel, $\zeta$ is the corresponding tuning bandwidth, and $\gamma=(1-\exp(-\frac{1}{2\zeta^2}))^{-1}$.\\
    
    $L_{sm}$ &$||\Delta G(v)||^2_2$ & The bias fields $G(v)$ estimated from the input intensity images should be smooth. A smoothness loss function is based on the Laplacian operator to provide explicit guidance of the smoothness constraint. \\
    
    $L_{ssim}$ & $-\log (\frac{1}{2|P|}\sum_{p\in P}(1 + SSIM(p)))$ & Structural similarity index loss. $P$ denotes the set of pixel location and $|P|$ is its cardinality.  $SSIM(\cdot)$ is one of the perceptual metrics and it is also differentiable.\\
    
    $L_{mc}$ & \makecell[l]{$E_{\sim\hat{x}_y}[-\log(D(\hat{x}_y, m_y))] +$ \\ $E_{\sim x}[-\log(D(x, m_x))]$} & Modality classification loss. Given the pair of input sample and target modality ${x, m_y}$, we propose to learn a parameterized mapping $f : {x, m_y} \rightarrow \hat{x}_y$ from ${x, m_y} $ to the generated corresponding sample with modality my to closely resemble $x_y$. $m_y$ denotes a four-dimensional one-hot vector to represent the four MR modalities\\
    
    $L_{gdl}$ & Gradient difference loss & To deal with the inherent blurry effect. \\
    $L_{edge}$ &  Edge-aware constraint & Based on the similarity of the edge maps from synthesized and real images.\\
    $L_{tumor}$ &  Tumor-aware constraint & The edge-aware learning helps the model capture the context information of the entire brain, but not necessarily for individual tumors. Since tumor appearance is highly variable and subject-specific, it is hard to synthesize than the normal tissue.\\
    $L_{mani}$ & Manifold regularizer & To preserve the complementary properties. Since the image manifold reflects the intrinsic geometric structure underlying the data leading to the generated results with a realistic overall structure.\\
    
    $L_{mmd}$ & Regularization &  Maximum mean discrepancy regularization \cite{gretton2012kernel}.\\
    $L_{gcr}$ &  Regularization & Geometry co-regularization. \\
    $L_{ae}$ & Reconstruction & Autoencoder reconstruction. \\
    $L_{oth}$ & Regularization & Spectral restricted isometry property regularization \cite{bansal2018can}.\\    
    $L_{fault}$ & Fault-aware discriminator & To make the synthesized results can satisfy the requirement of segmentation and substitute the real acquisitions in practice, and bridge the gap of segmentation performance between synthesized data and real ones.\\
    \hline
    \end{tabular}
}
\label{tab:summary-loss-second}
\end{table*}

\subsection{Losses}\label{sec:losses}
Table~\ref{tab:summary-loss-first} and Table~\ref{tab:summary-loss-second} provide a comprehensive overview of the loss functions in cross-modality neuroimage synthesis. The second column in Table~\ref{tab:summary-loss-first} and Table~\ref{tab:summary-loss-second} denotes the loss function. The third column describes the details or the role of the loss function. \revised{We categorize the overall loss functions into three streams. The first category is the basic class of GAN loss functions, including the adversarial loss function $L_{gan}$ and its variants $L_{sgan}$ and $L_{cgan}$, the cycle loss function $L_{cyc}$, the pixel identity loss function $L_{p2p}$, the edge identity loss function $L_{edge}$. They are commonly used for the general cross-modality synthesis but are not specifically designed for cross-modality neuroimage synthesis. The second category of loss functions is the regularizer loss function. They construct the optimization problems by utilizing the intrinsic characteristic of neuroimaging. For instance, $L_{mmd}$ is the maximum discrepancy regularization when seeking the paired feature points from the source and target modality neuroimage. $L_{cor}$ stabilizes the training procedure and improves the bias field quality by suppressing the negative influence from the outlier noise. $L_{tumor}$ utilizes the similarity contextual information for both T1 and T2 modalities to regularize the quality of synthesized neuroimage. $L_{sc}$ regularizes the generator to ensure adherence to the consistency of tumor shape among multi-modalities neuroimages. The third category of loss functions is strongly related to the downstream task, like segmentation loss $L_{seg}$, dice score loss $L_{dice}$, cross-entropy loss $L_{cls}$, and modality classification loss $L_{mc}$. These loss functions aim to optimize the downstream task with cross-modality neuroimage synthesis jointly.}

\begin{table*}[ht]
\caption{A brief summary of different metrics for cross-modality neuroimage synthesis.}
\renewcommand{\arraystretch}{1.2}
\resizebox{\textwidth}{!}{
    \begin{tabular}{llp{5cm}p{13cm}}
    \hline
    \textbf{Acronyms} & \textbf{Level} & \textbf{Name} &  \textbf{Description} \\\hline
    PSNR & $\uparrow$ & Peak signal-to-noise ratio & \\
    
    SSIM & $\uparrow$ & Structural similarity &\\
    MSSSIM & $\uparrow$ & Multiscale structural similarity &\\
    
    MAE & $\downarrow$ & Mean absolute error  & \\
    
    DSC & $\uparrow$ & Dice similarity coefficient & DSC is employed to evaluate different approaches by comparing their segmentation results against the ground truth voxel-by-voxel. \\
    
    ASD/ASSD & $\uparrow$ & Average (symmetric) surface distance & It is used to compute the average surface distance from $y_{pred}$ to $y$ under the default setting. This tells us how much, on average, the segmentation surface deviates from the GT.\\
    
    MOS & $\uparrow$ & Mean opinion score & Measure the quality of a given image by a rating score between 1 and 5: 1 indicates inferior while 5 indicates superior. \\
    
    DS & $\downarrow$ & Deformation score & A registration metric based on Jacobian. For DS, an FFD-based \cite{rueckert1999nonrigid} deformable registration is applied to the synthesised MR to register it to a real MR at a similar imaging plane. The average Jacobian (normalised to [0,1]) of the required deformation to complete such registration was computed as the score consequently. The underlying assumption is that a synthesised MRI with high quality tends to have a lower Jacobian when registering to the real MRI.\\
    P & $\uparrow$ & Precision &  The fraction of true positive examples among the examples that the model classified as positive. \\
    R & $\uparrow$ & Recall & The fraction of examples classified as positive, among the total number of positive examples. \\
    F1S & $\uparrow$ & F1 score for classification & It is defined as $2(P \times R)/(P + R)$. \\
    F-score  & $\uparrow$ & F1 score for segmentation & Measure the overlap of ground truth segmentation labels. It is defined as $(2|H \cap G|)/(|H|+|G|)$ where $G$ is the label of the target image and $H$ is the prediction of the source image.\\
    VS & $\uparrow$ & Verisimilitude score & A higher VS means more produced pseudo-healthy images fool the classifier successfully, indicating the model can better generate lesion-free images lying in the true data distribution of normal images. \\
    
    MSE & $\downarrow$ & Mean-squared error & \\
    NMSE & $\downarrow$ & Normalized mean-squared error & \\
    RMSE & $\downarrow$ & Root mean-squared error & \\
    NRMSE & $\downarrow$ & Normalized root mean-squared error & \\
    
    HD95 &  $\downarrow$ & 95th Hausdorff Distance &  \\
    CC & $\downarrow$ & Correlation coefficient  & \\
    CIV & $\uparrow$ & Coefficient of intensity variation  & \\
    L1 & $\downarrow$ & L1 error  &  \\
    L2 & $\downarrow$ & L2 error &  \\
    t-L2 & $\downarrow$ &   Tumor-averaged L2 & Based on segmentation maps of enhancing lesions.\\
    FID & $\downarrow$ & Fréchet inception distance &  For the evaluation of the performance of synthesis models at image generation \cite{heusel2017gans}. \\
    
    UQI & $\uparrow$ & Universal quality index & This index is designed by modeling any image distortion as a combination of three factors: loss of correlation, luminance distortion, and contrast distortion \cite{wang2002universal}.\\
    
    MMD & $\downarrow$ & Maximum mean discrepancy &  \\

    IS & $\uparrow$   & Inception score & \\
    ACC & $\uparrow$  & Accuracy & \\
    AUC & $\uparrow$  & Area under curve & Area under the receiver operating characteristic.\\
    ROC & $\uparrow$  & Receiver operator characteristic & Show the diagnostic ability of binary classifiers. \\
    BAC & $\uparrow$  & Balanced accuracy & \\
    SEN & $\uparrow$  & Sensitivity & \\
    SPE & $\uparrow$  & Specificity & \\
    MCC & $\uparrow$  & Matthews correlation coefficient & \cite{matthews1975comparison}\\
    DVR & $\uparrow$  & [$^{11}$C]PIB PET distribution volume ratio & Relect the myelin content. \\
    MI & $\uparrow$ & Mutual information & \cite{wang2004image} \\
    Q & $\uparrow$ & Weighted fusion quality metric & \cite{piella2003new} \\
    FMI  & $\uparrow$ & Feature mutual information measures & \cite{haghighat2014fast}\\
    
    \hline
    \end{tabular}
}
\label{tab:summary-metric}
\end{table*}

\subsection{Metrics}\label{sec:metrics}
Table~\ref{tab:summary-metric} provides a comprehensive review of the metrics in cross-modality neuroimage synthesis. The first and third columns list the acronym and full name of a metric, respectively. The up arrow in the second column indicates that a higher value represents a better result; the opposite applies for the down arrow. The last column provides additional descriptions, which are tailored for cross-modality neuroimage synthesis if possible. From Table~\ref{tab:summary-metric}, it can be easily observed that most of the novel metrics are variants of traditional metrics, such as MAE, PSNR, or SSIM. \revised{The traditional metrics are fast to compute; however, they cannot reflect the true quality of neuroimage because they focus more on the low-level pixel-wise fidelity and ignore the fundamental traits of neuroimages, such as the structural characteristics of the lesion area and the k-space feature drift. One alternative approach to alleviate this problem is to use downstream tasks such as segmentation or classification to validate the quality of generated samples~\cite{Huang2017SimultaneousSA,Huang2018CrossModalityIS}. Nevertheless, The drawbacks associated with this approach are as follows. 1) The computational expense is substantial, often necessitating the utilization of GPUs for computation. 2) The additional segmentation significantly influences the inference speed. 3) This approach cannot be seamlessly integrated into an end-to-end optimization for cross-modality neuroimage synthesis. Hiring a radiologist for review is another option, but it is too costly, time-consuming, and difficult to scale up. Hence, the validity of these metrics for neuroimages remains to be explored.}



\xgy{\section{Open challenges and future directions}\label{sec:future_work}}

\revised{As an emerging area, research on multi-modality brain image synthesis is still in its infancy. The remaining challenging topics and potential solutions are summarized as follows. 

\textbf{\textit{Challenge 1: Joint optimization of the synthesis model and downstream task.}} How to jointly optimize the cross-modality neuroimage synthesis and the downstream tasks with either weakly supervised or unsupervised learning? Previously, neuroimage synthesis~\cite{Huo2019SynSegNetSS,Shin2020GANDALFGA} is generally regarded as a standalone task, which overlooks one crucial fact, namely whether the synthesized results can improve the downstream tasks. 
    
\textit{\textbf{Solution 1: Multi-task learning.}} Transforming downstream tasks and cross-modality synthesis into multi-task learning improves both the quality of synthesized neuroimages and the performance of downstream tasks. Recently, there are some initial attempts in the joint optimization of cross-modality neuroimage synthesis and downstream tasks. Unsupervised learning methods~\cite{Yu2021MouseGANGM,Sun2020AnAL,He2021AutoencoderBS} and weakly supervised learning approaches~\cite{Shen2021MultiDomainIC,Zhou2021AnatomyConstrainedCL} start paying attention to downstream tasks rather than focusing only on the quality of the synthesized results. 

\textbf{\textit{Challenge 2: Correctness of synthesized lesions.}} How to ensure the correctness of the synthesized lesions? Previously, most cross-modality image synthesis algorithms pay attention to the quality of the whole image, which fail to highlight the essential regions related to disease (see Fig.~\ref{fig:existing_dismerit}).

\textit{\textbf{Solution 2: Lesion-oriented synthesis.}} The method reported in \cite{Sun2020AnAL} provides a potential answer to this question since the diagnosis of the lesion can be treated as an anomaly detection task. If we can detect the disease region and use it as guidance for cross-modality brain image synthesis, the generated output is steered in a disease-highlighted and lesion-oriented manner.

\textbf{\textit{Challenge 3: Dedicated evaluation metrics.}}  How to identify an appropriate metric to evaluate the results of cross-modality image synthesis? Existing measurements are usually based on the PSNR and SSIM, which are established on natural images but ignore the inherent properties of neuroimages. The translated medical data with the highest PSNR or SSIM may still be blurred, or miss important neural tissue representations.

\textit{\textbf{Solution 3: K-space-aware metrics.}} The concept of k-space imaging is the main distinction between MR imaging and natural images, in addition to the disease area. The new synthetic quality metric should accurately reflect the location of the critical lesion and the essential k-space characteristics of neuroimages. Therefore, the new metric should encompass the lesion region, k-space characteristics, and anatomical features.

\textbf{\textit{Challenge 4: Misaligned data usage.}} How to utilize misaligned or unpaired neuroimaging data for cross-modality image synthesis? In practice, a large amount of misaligned neuroimaging data exists in each vendor (i.e., hospital). The state-of-the-art image registration algorithm takes a lot of time to register a pair of brain images, especially across modalities. It also requires a great deal of laborious work to verify the registration result, and correct residual errors if necessary. Whether the strong dependence on registration can be eliminated for cross-modality synthesis is questionable.

\textit{\textbf{Solution 4: Misaligned neuroimage as data augmentation.}} Kong~\textit{et al.}~\cite{Kong2021BreakingTD} and Wang~\textit{et al.}~\cite{Wang2022FedMedATLMU} attempt to eliminate the need for registration and make full use of misaligned neuroimaging data for synthesis. The work of~\cite{Kong2021BreakingTD} incorporates the correction loss into CycleGAN~\cite{Zhu2017UnpairedIT}, while Wang~\textit{et al.}~\cite{Wang2022FedMedATLMU} regard the misaligned neuroimaging data as a data augmentation of the self-supervised learning method and design an affined transformation loss to enable the discriminator overcoming the over-fitting problem. Furthermore, the authors in~\cite{Wang2022FedMedATLMU} stimulate severely misaligned neuroimaging data and find that their methods perform better in the presence of severe misalignment.

\textbf{\textit{Challenge 5: A unified model.}} How to build up a unified model for cross-modality brain image synthesis? 

\textit{\textbf{Solution 5: One to many via invariant features.}}  
    Previously, most works focused on modality synthesis in MR, CT, and PET. Charts~\textit{et al.}~\cite{Chartsias2018MultimodalMS} propose a multi-input and multi-output fully convolutional network model to synthesize various modalities of MRI. Similarly to the work of~\cite{Chartsias2018MultimodalMS}, Liu~\textit{et al.}~\cite{Liu2021AUC} propose a unified conditional disentanglement work to synthesize various modalities of MRI. \zz{Moreover, in order to speed up the inference phase, Zhang~\textit{et al.}~\cite{zhang2020inductive} provides an excellent way to speed up the modal-specific feature retrieval by transferring knowledge from original data to hash codes}. However, simultaneous synthesis of varying MR, CT, and PET modalities is lacking. The work of~\cite{Zhou2021SynthesizingMP} may provide a solution, which uses a cycle-consistent GAN to extract the invariant features from various modalities. Nevertheless, its synthesis mode is limited in MRI-to-MRI. The invariant characteristics from CT, PET, and MRI should be extracted to train a unified synthesis model in the future.

\textbf{\textit{Challenge 6: Data privacy.}} How can the data isolation problems be solved while protecting the patients' privacy for cross-modality brain image synthesis? The current state-of-the-art brain image synthesis algorithms consider a centralized training strategy. However, many medical institutions cannot share their data, which is restricted by privacy protection legislation.

\textit{\textbf{Solution 6: Federated cross-modality neuroimage synthesis.}} Wang~\textit{et al.}~\cite{wang2023fedmed} are the first to address this issue. They find out that the clipped gradient enhances the data privacy guarantee and increases the fidelity of synthesized neuroimages. Then FedMed-ATL~\cite{Wang2022FedMedATLMU} presents a more unbalanced data setting for clients to simulate the more realistic scenario.
}
\minor{\section{Future Practical Applications}\label{sec:pratical_application}
There are two scenarios in which cross-modality neuroimage synthesis can be applied. First and foremost, as mentioned in \textbf{\textit{Challenge 1}} of Section~\ref{sec:future_work}, cross-modality neuroimage synthesis serves as an auxiliary method for diagnosis, such as investigating pathology and neurodegeneration~\cite{Pan2021DiseaseimagespecificLF}. Therefore, it is necessary to ensure that the synthesized neuroimage is optimized to improve the performance of downstream tasks, such as tumor detection~\cite{Yu2021MouseGANGM,Sun2020AnAL,He2021AutoencoderBS} and segmentation~\cite{Shen2021MultiDomainIC}. To achieve this goal, utilizing multi-task learning to simultaneously optimize the synthesis of neuroimages across different modalities and their downstream tasks is an up-and-coming area of research. Second, as discussed in \textbf{\textit{Challenge 4}} of Section~\ref{sec:future_work}, using cross-modality neuroimage synthesis to minimize the manual effort required for registration is an additional area of great potential. Previously, the registration process for misaligned neuroimages required a significant amount of manual effort~\cite{avants2011reproducible,klein2009evaluation}, leading to low throughput in practice. Registration-free cross-modality neuroimage synthesis gains attraction in academia~\cite{chen2020reusing,Kong2021BreakingTD}, which can dramatically reduce physicians' workload. However, physicians still need to verify the content generated, and further future exploration of its interpretability is required. Third, as mentioned in \textbf{\textit{Challenge 5}} of Section~\ref{sec:future_work}, there is a scarcity of cross-modality neuroimage synthesis models that can simultaneously synthesize MR, CT, and PET. Deploying multi-synthesis models in medical institutions is impractical due to limited storage capacity. Therefore, developing a cohesive cross-modality neuroimage synthesis model within the medical imaging community is imperative. Finally, as presented \textbf{\textit{Challenge 6}} of Section~\ref{sec:future_work}, data privacy remains a significant concern in the field of cross-modality neuroimage synthesis. Training a cross-modality neuroimage synthesis model often requires a large number of neuroimages. However, collecting a large number of neuroimages for training is challenging due to privacy regulations. Hence, incorporating federated learning into cross-modality neuroimage synthesis holds great potential in effectively leveraging a multitude of neuroimages from different medical institutions while adhering to privacy regulations. }
\section{Conclusion}

In this paper, we provided a literature review on cross-modality brain image synthesis, focusing on the level of supervision, downstream tasks, modalities, datasets, loss functions, and evaluation metrics. In particular, we characterized the architecture of existing cross-modality synthesis models based on the supervision level. In addition, for each downstream task, we analyzed in detail how cross-modality neuroimage synthesis enhances its performance. In the end, we highlighted a number of exciting future research directions for cross-modality neuroimage synthesis. \revised{In conclusion, the recent prevalence of cross-modality neuroimage synthesis significantly impacted downstream tasks in the medical imaging field. However, multi-task learning with cross-modality neuroimage synthesis and other downstream tasks were still in the early research phase. We hope the review and analysis presented in this paper will inspire researchers to develop more effective cross-modality neuroimage synthesis methods for practical applications.}

\section*{Acknowledgments} 
This work is partially supported by the National Key R\&D Program of China (Grant NO. 2022YFF1202903) and the National Natural Science Foundation of China (Grant NO. 62122035 and 62206122). Y. Jin is supported by an Alexander von Humboldt Professorship for AI endowed by the German Federal Ministry of Education and Research.


\bibliographystyle{ACM-Reference-Format}
\bibliography{sample-base}

\end{document}